\documentclass{IEEEtaes}
\usepackage{cite}
\usepackage{graphicx}
\usepackage{epstopdf}
\usepackage{booktabs}
\usepackage{float}
\usepackage{algorithm}
\usepackage{algpseudocode}
\usepackage{amsmath}
\usepackage{mathrsfs}
\usepackage{amsthm}
\usepackage{amsfonts}
\usepackage{amssymb}
\usepackage{bm}
\usepackage{arydshln}
\usepackage{color}
\usepackage{caption}
\usepackage{colortbl}
\usepackage{mathtools}

\renewcommand\d{\displaystyle}

\usepackage{color,array,amsthm}
\usepackage{graphicx}

\newtheorem{assumption}{Assumption}
\jvol{XX}
\jnum{XX}

\newtheorem{theorem}{Theorem}
\newtheorem{lemma}{Lemma}
\begin{document}

\title{Coordinated  Control of Deformation and 
Flight for Morphing Aircraft via Meta-Learning and Coupled State-Dependent Riccati Equations}

\author{HAO-CHI CHE}
\member{}
\affil{Beihang University, Beijing, China} 

\author{HUAI-NING WU}
\affil{Beihang University, Beijing, China\\ Hangzhou International Innovation Institute, Beihang University, Hangzhou, China} 


\receiveddate{This work was supported in part by the National Natural Science Foundation of China under Grants 92271115 and 62473021, and in part by the Research Start-up Funds of Hangzhou International Innovation Institute of Beihang University under Grant 2024KQ059.}

\corresp{ (Corresponding author: Huai-Ning Wu)}

\authoraddress{Hao-Chi Che is with the School of Automation Science and Electrical Engineering, Beihang University, Beijing, 100191, China 
(e-mail: chehaochi@126.com). Huai-Ning Wu is with the  School of Automation Science and Electrical Engineering, Beihang University, Beijing, 100191, China, and also with Hangzhou International Innovation Institute, Beihang University, Hangzhou, 311115, China (e-mail: whn@buaa.edu.cn).}

\markboth{CHE ET AL.}{CCDFMA}
\maketitle

\begin{abstract}In this paper, the coordinated control problem of deformation and flight for morphing aircraft (MA) is studied by using meta-learning (ML) and coupled state-dependent Riccati equations (CSDREs). 
Our method is  built  on two principal observations that dynamic models of MA  under varying morphing conditions share a morphing
condition independent representation function and that the specific morphing condition part lies in a set of linear coefficients. 
To that end, the domain adversarially invariant meta-learning (DAIML) is employed to learn the shared representation with offline flight data. Based on the learned representation function, the coordinated control of the deformation and flight for MA is formulated as a non-cooperative differential game. The Nash equilibrium strategy
can be derived by addressing a pair of CSDREs. For this purpose,
 Lyapunov iterations are extended to obtain the  stabilizing solutions 
of the CSDREs,  and the convergence proof of the proposed algorithm is provided. 
Finally, a simulation study is carried out to validate the efficacy of the developed coordinated game control strategies.
\end{abstract}


\section{INTRODUCTION}

 The morphing aircraft (MA) can change the geometric shape actively or passively in flight 
in response to the varying environments and missions, to guarantee the optimal flight performance 
in various flight stages.  In this case, deformable flight technology has become  a focal point of research among countries around the world  \cite{Bonnema1988AFTIF111MA,Pendleton2000ActiveAW,Kudva2004OverviewOT}.

The ability to morph, while bringing performance improvements to the aircraft, also presents new requirements and challenges for its decision and control.  Currently, the ways of designing morphing control are  categorized into two  types: the first is  independent  of flight control design while
the second is coordinated with flight control design\cite{Seigler2007ModelingAF}. 
In previous studies, most scholars focused on the former, wherein the deformation of MA was determined by  predetermined program and rates. The significant fluctuations in mass distribution and aerodynamic
forces resulting from deformation were  considered as perturbations. Flight controller was employed to stabilize the transition between various configurations. 
With the development of extended state observer and disturbance observer \cite{boutayeb1999strong} and their successful application in nonlinear control systems, some scholars have introduced these methods into the control field of MA to suppress disturbances and ensure the stable flight of aircraft \cite{gong2019disturbance,wu2019new}. 
In addition, to guarantee stability during the transition between various configurations,   the design of MA attitude controller  incorporated switching control \cite{JIANG20151640}, linear variable parameter control \cite{lee2019linear}, and gain schedule $H_\infty$ control methods \cite{YUE2013909}.  
On the {basis} of Lyapunov stability theory \cite{seigler2009analysis} and Hurwitz stability criterion \cite{shi2015morphing}, the stability conditions during the transition of MA were obtained.

The aforementioned studies primarily focused on designing controllers for the time-varying parameters and additional disturbances introduced by the deformation process, without exploring the variable wing as an active control method. 
Most of the related works were achieved by considering structural deformation as an auxiliary manipulation method, and the integration of morphing wings with conventional aerodynamic actuators can  assist in improving the maneuverability of aircraft \cite{gandhi2009hardware, guo2018compound,yin2015coordinated}. 
Nevertheless, all the aforementioned results require precise knowledge of the relationship between the morphing parameters and the model.  
Some researchers used methods such as   approximate dynamic programming \cite{liu2020optimal} to design optimal morphing strategies.
  It is important to highlight that  above algorithm is limited to only a few fixed deformation positions, whereas in actual flight scenarios, the morphing control is usually continuous. 

Using a model for  optimal control  enables faster acquisition of optimal strategies. Nevertheless, the necessity for a model may  impose limitations on both the  robustness and applicability. 
 Recently, artificial intelligence technologies such as deep learning \cite{lecun2015deep}, reinforcement learning (RL) \cite{littman2015reinforcement} and deep RL \cite{mnih2015human,lillicrap2015continuous,schulman2017proximal} are  developing rapidly, and  providing new ideas for intelligent morphing decision and flight control of MA \cite{valasek2005reinforcement,wen2017deep,li2020morphing}. 
Machine learning techniques (e.g., deep neural networks (DNNs)) possess  high representation power but are often too slow to update onboard. Hence, many existing learning methods are unable to estimate the complex dynamics presented by MA flying at different morphing conditions for real-time.
Recently, scholars have been addressing the data and computation demands associated with  DNNs to advance the field towards the fast online-learning paradigm, with the rising interest
 in few-shot learning \cite{snell2017prototypical}, continual learning \cite{kirkpatrick2017overcoming,zenke2017continual}, and meta-learning (ML) \cite{santoro2016meta}.
The most prevail learning framework utilized in dynamic environments is ML, or “learning to learn”. This approach seeks to develop an efficient model by leveraging data from various tasks or environments \cite{finn2017model,hospedales2021meta}. 
For MA applications, the flight of MA under different morphing conditions can correspond to different tasks. The goal of ML is to learn a common representation that can adapt to different morphing conditions. The ability of ML to “learning to learn", i.e., it can rapidly learn new tasks by leveraging pre-existing “knowledge", makes it possible to quickly adjust the deformation of MA models.  

{The various subsystems of a MA (such as the structural deformation module, attitude control system, propulsion system, etc.) often have different optimization objectives. 
Additionally, changes in aerodynamic parameters caused by structural deformation may interfere with flight control, potentially leading to conflicts of interest. The non-cooperative game allows each subsystem to independently optimize its own objectives while achieving equilibrium through strategic interactions, thus avoiding performance losses caused by forced cooperation (unified goals). Therefore, the interaction between morphing control and flight control can be viewed as two independent agents responding to specific flight tasks using different means based on their own interests. In our work, we utilize the non-cooperative game framework to construct an interaction model for deformation control and flight control, ultimately reaching a Nash equilibrium state.} Based on the common representation obtained through ML,   the coupled state-dependent Riccati equation (CSDRE) method is employed to derive  a Nash equilibrium strategy. 
Since the mid-1990s, state-dependent Riccati equation (SDRE) control has been recognized as a viable alternative approach for nonlinear control design \cite{cloutier1996nonlinear,cloutier1997state,dutka2005optimized}. A comprehensive overview of the recent advancements in SDRE approach has been presented by $\c{C}$imen in  \cite{2012Survey}. 
Inspired by the SDRE framework, 
the study  in \cite{wang2018coupled} considers the CSDRE approach for systematic design of nonlinear quadratic regulator and design  of $H_\infty$ control for mechatronics systems. Nevertheless, how to solve these CSDREs becomes a new challenge. Currently,  there exists no universally effective approach for solving these CSDREs.

Motivated by aforementioned works, the coordinated control problem of deformation and flight for MA is studied by using ML and CSDREs. Our  approach comprises two primary parts (depicted in Fig. 1): an offline learning stage and an online game stage. During the offline learning stage, the domain adversarially invariant meta-learning (DAIML) is employed to learn a morphing condition independent   representation function
of unknown dynamics in a data-efficient way. 
In the online game stage, a non-cooperative differential game is constructed to describe the coordinated control of the deformation and flight for the MA with the learned representation as a basis.   

The primary innovations and benefits of the proposed method, in comparison to current results, are outlined as follows:

\begin{enumerate}
\vspace{-6pt}
\item{\emph{DNN approximating unknown models:}} Compared with the results in \cite{guo2018compound} and \cite{yin2015coordinated} which require  the relationship between the morphing ratio and the  model parameters to be accurately known,  we employ a DNN to approximate the  model dynamics of MA. To help with training, DAIML technique is used to optimize the network useful for subsequent game phase.

\item  {\emph{Coordinated control for MA:}} Different from the results in \cite{lee2019linear,JIANG20151640,YUE2013909,seigler2009analysis,shi2015morphing}  where structural deformation was directly applied to the MA as an external command,  this  paper presents a non-cooperative game  to achieve coordinated  operation of structural deformation and flight maneuvering for MA.  
Unlike the result in \cite{liu2020optimal},  the optimal morphing strategies we obtained are no longer a few fixed shapes.

\item {\emph{CSDREs solving method based on Lyapunov iterations:}} {Compared to the results in \cite{wang2018coupled},  an algorithm based on Lyapunov iterations  is extended to calculate a pair of stabilizing solutions of the CSDREs. Moreover, the proof of the algorithm's  convergence is provided.}
\vspace{-6pt}
\end{enumerate}

The paper is structured as follows: Section II gives the preliminaries and the problem formulation of this study.  
After that, in Section III, the DAIML is extended to learn a morphing condition independent DNN representation for approximating the unknown dynamics of MA. Additionally, a classifier is trained to represent the mapping relationship between states and morphing ratio. 
In Section IV, the differential game model for  coordinated control of deformation and flight is constructed and an algorithm to calculate a pair of stabilizing solutions of the CSDREs  is proposed. Moreover, the detailed steps to implementation of CSDRE controller are provided. Finally, the effectiveness of the proposed strategies is illustrated by numerical simulations in Section V.
\begin{figure}[h]
	\centering
	\includegraphics[width=8cm, height=7cm]{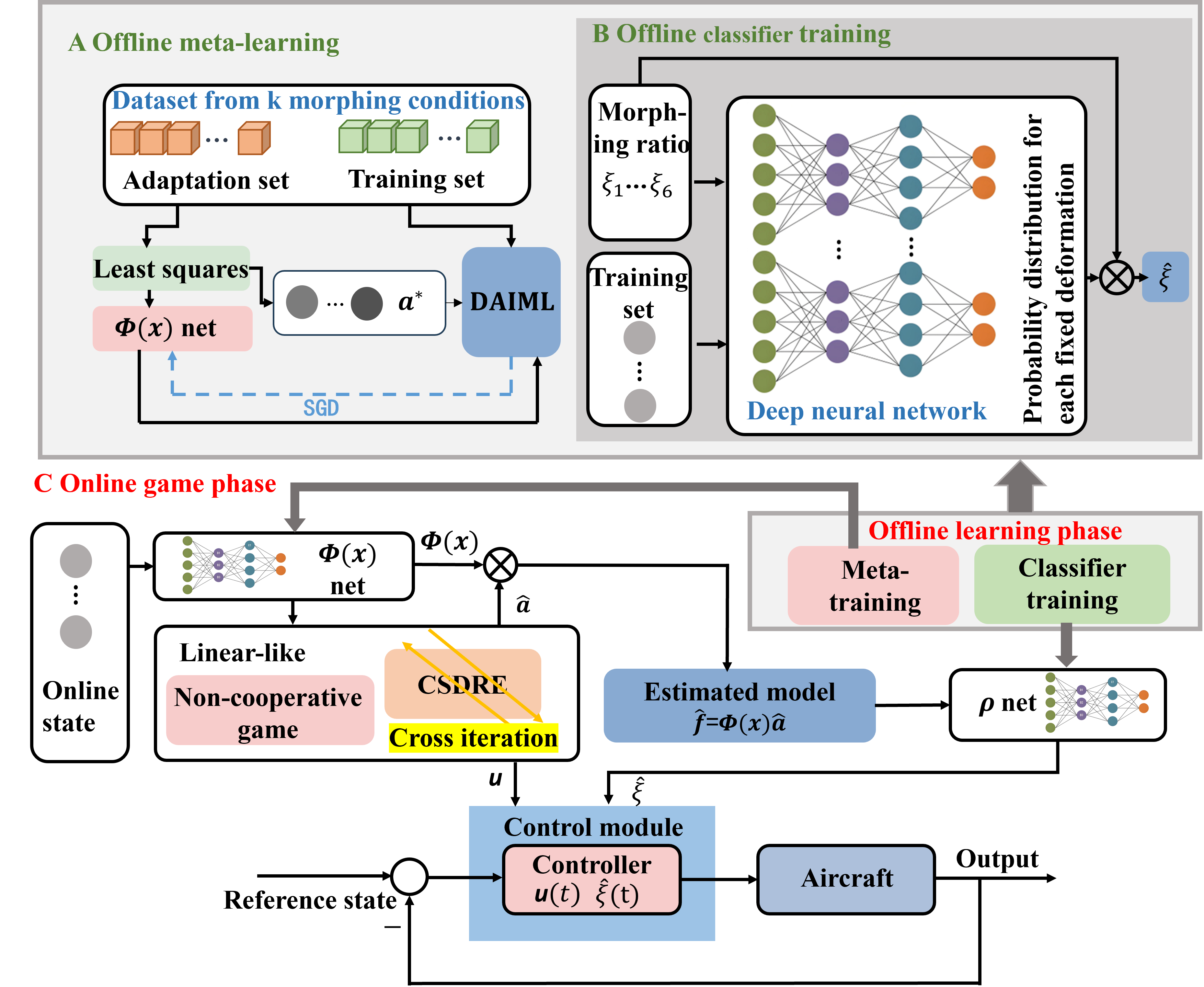}
\caption{{\small $\mathbf{Offline~learning~phase~and~online~game~pahse}$. ($\mathbf{A}$) This part provides an explanation of the ML algorithm DAIML.We collect data from $\{\xi_1,~\xi_2,~,\cdots,\xi_6\}$ under different deformation conditions and apply DAIML to train the $\bm\Phi$ network. ($\mathbf{B}$) This part represents the classifier training module.  A mapping relationship between the model and wingspan morphing ratio is trained based on the DNN. ($\mathbf{C}$) The diagram shows our design process, with the gray module representing the offline learning phase. On the
basis of learned network $\bm\Phi$, a linear-like model is constructed, and the deformation specific weight $\hat{\bm a}$ and flight control strategy $\bm u$ are updated using a non-cooperative game framework.}}
\label{fig8}
\vspace{-15pt}
\end{figure}

$Notations:$ 
{\small$||\cdot||$} indicates the vector norm or matrix norm. 
{\small$\otimes$} denotes Kronecker product. For a matrix {\small$ \bm Q$},  {\small${\rm vec}(\bm Q)$} denotes the vectorization of  matrix {\small$\bm Q$}, and  {\small${\rm sym}(\bm Q)$} is  defined as {\small$ \bm Q+\bm Q^{T}$}. For a matrix {\small$\bm M$} and a vector {\small$\bm v$}, {\small$||\bm v||_{\bm M}^{2}\triangleq \bm v^{T}\bm M \bm v$}.  {\small$\Omega$} is defined as   a bounded open subset of
some Euclidean space, such that  {\small$\bm 0 \in \Omega\subseteq \mathbb{R}^{n}$}.    A function is  classified as belonging to the class {\small$\mathcal{C}^{k}(\Omega)$} (or simply {\small$\mathcal{C}^{k}$})  if it is continuously differentiable {\small$k$} times within the domain {\small$\Omega$}. The notation {\small$\mathcal{C}(\Omega)$} (or {\small$\mathcal{C}^{0}$}) specifically refers to the set of continuous functions defined in {\small$\Omega$}.
\vspace*{-4mm}
\section{ PRELIMINARIES AND PROBLEM STATEMENT}
\subsection{Model of  MA}
\vspace{-5pt}
\begin{figure}[h]	
\centering
	\includegraphics[width=7cm, height=3cm]{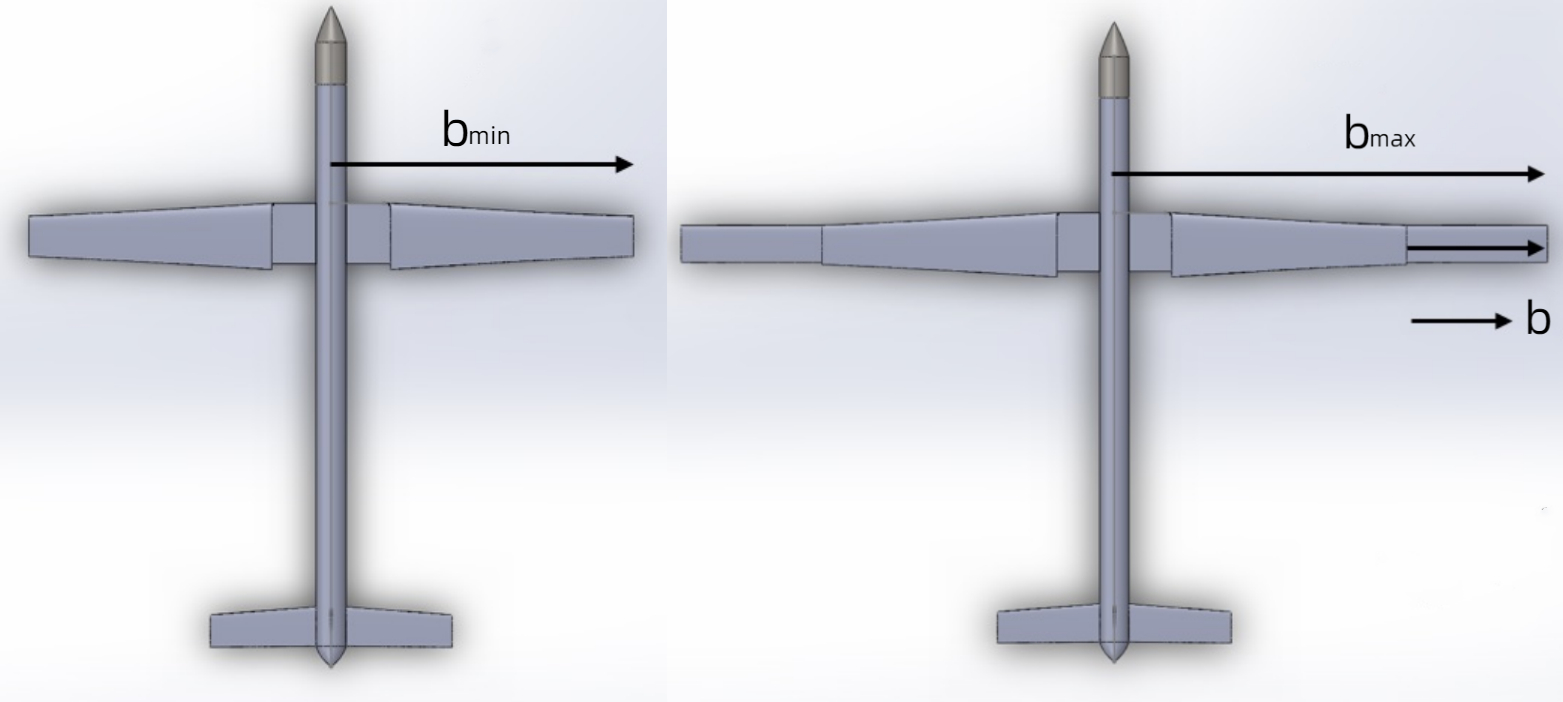}
	\caption{{\small$\mathbf{Variable}$-$\mathbf{span~MA. }$}}
	\label{fig8}
\end{figure}
\vspace{-15pt}
As illustrated in Fig. 2, the MA model in \cite{yin2015coordinated} is used in this
study. The morphing ratio is defined as
{\small$
	\small\d\xi_n=\frac{b-b_{\min}}{b_{\max}-b_{\min}},
$}
where, {\small$ b$} is the wingspan, {\small$b_{\min}$} and {\small$b_{\max}$} are the shortest and longest wingspans, respectively. Obviously, {\small$\xi_n\in[0,~1]$}.

The  longitudinal nonlinear dynamic model for a MA can be represented as follows: 
\vspace{-6pt}
\begin{equation}\label{equ:C1}	
\small\left\{\begin{split}
		&\dot{V}=\frac{1}{m}T\cos\alpha-\frac{1}{m}D(\xi_n)-g\sin(\theta-\alpha)\\
		&\dot{\alpha}=-\frac{1}{mV}T\sin\alpha-\frac{1}{mV}L({\xi_n})+q+\frac{1}{V}g\cos(\theta-\alpha)\\
		&\dot{\theta}=q\\
		&\dot{q}=\frac{1}{I_y}M({\xi_n})\\
		&\dot{h}=V\sin(\theta-\alpha)\\
	\end{split}\right.
\end{equation}
where {\small$\alpha$}, {\small$\theta$} and {\small$q$}  are  attack angle, pitch angle and  pitch angular velocity, respectively, {\small$V$} denotes the flight velocity, {\small$h$} is the flight altitude, {\small$I_y$}
is pitch  moment of inertia. {\small$T$} is thrust and 
{\small$T=T_{\delta_t}\delta_t$},
where {\small$T_{\delta_t}$} is the thrust coefficient and {\small$\delta_t$} is the throttle opening. 

The specific forms for aerodynamic force and aerodynamic moment are:
\vspace{-3pt}
\begin{equation}\label{equ:C2}
\small\left\{\begin{split}
		&L({\xi_n})=QS_wC_L\\
		&D({\xi_n})=QS_wC_D\\
		&M({\xi_n})=QS_wc_AC_M
	\end{split}\right.
\end{equation}
where {\small$S_w$}  and  {\small$c_A$} are   reference area and    mean aerodynamic chord, respectively,  {\small$Q=\d\frac{1}{2}\rho V^2$},   {\small$\rho$} and {\small$Ma$} are dynamic pressure, air density and  Mach number, respectively.  {\small$Q$} and  {\small$Ma$} depend on the {\small$\rho$} and sound speed {\small$\mathfrak{o}$} at the current flight altitude {\small$h$}. Within the troposphere (below an altitude of about $10$ km), these parameters adhere to  the fitting calculation formula {\small$\d\rho = 1.2250(1-\frac{h}{44.3308})^{4.2559}$}  and {\small$Ma=\d\frac{V}{\mathfrak{o}}$}, where {\small$\mathfrak{a} = 20.0468\sqrt{288.15-6.5h}$} \cite{Yang1983}.
{\small$C_L$}, {\small$C_D$} and {\small$C_M$} are the aerodynamic parameter functions related to wingspan variable, which are
influenced by morphing ratio {\small$\xi$}. 
Within the normal  attack angle, they can be approximated as
\vspace{-3pt}
\begin{equation} \label{equ:C3}	
\small\left\{\begin{split}
		C_L=&C_{L_{\alpha=0}}(Ma,\xi_n)+C_{L_{\alpha}}(h,Ma,\xi_n)\alpha\\&+C_{L_{\delta_e}}(h,Ma)\delta_e+C_{L_{q}}(Ma,\xi_n)\frac{c_A}{2V}q\\
		C_D=&C_{D_{\alpha=0}}(h,Ma,\xi_n)+C_{D_{\alpha}}(h,Ma,\xi_n)\alpha\\&+C_{D_{\alpha^2}}(h,Ma,\xi_n)\alpha^2\\
		C_M=&C_{M_{\alpha=0}}(h,Ma,\xi_n)+C_{M_{\alpha}}(h,Ma,\xi_n)\alpha\\&+C_{M_{\delta_e}}(h,Ma)\delta_e+C_{M_{q}}(Ma,\xi_n)\frac{c_A}{2V}q
	\end{split}\right.
\end{equation}
where {\small$\delta_e$} is elevator deflection angle, {\small$C_{L_{\alpha=0}}$}, 
 ..., {\small$C_{M_{q}}$} are the corresponding longitudinal aerodynamic parameters, respectively.\footnote{The longitudinal aerodynamic parameters in \eqref{equ:C3} are functions of {\small$h$}, {\small$Ma$}, and {\small$\xi_n$}. It can be obtained through least squares (LS) fitting using simulation software \cite{yin2015coordinated}. In this work, the machine learning techniques are utilized to approximate the unmodeled
	dynamics term, thereby eliminating the necessity of using LS  fitting.}

By integrating \eqref{equ:C1}-\eqref{equ:C3}, the affine nonlinear model of the MA can be derived as follows:
\vspace{-6pt}
\begin{equation}\label{equ:C5}	
\small\begin{split}
		\dot{\bm{x}}_n=\bm{f}_n(\bm{x}_n,\xi_n)+\bm{g}_n(\bm{x}_n)\bm{u}_n\\
	\end{split}
\end{equation}
where {\small$\bm{x}_n=[V~~\alpha~~\theta~~q~~h]\in{ \mathbb{R}^5}$}  denotes the state vector, and {\small$\bm{u}_n=[\delta_e~~\delta_t]\in{ \mathbb{R}^2}$} is the control input vector, {\small$\bm{f}_n(\bm{x}_n,\xi_n)$} and {\small$\bm{g}_n\big(\bm{x}_n\big)$} are
given by
\begin{equation*}
	\small\begin{split}
		&\bm{f}_n\big(\bm{x}_n,{\xi_n}\big)=\\
		&\begin{bmatrix}
			-\d\frac{1}{2m}\rho V^2S_w(C_{D_{\alpha=0}}+C_{D_{\alpha}}\alpha+C_{D_{\alpha^2}}\alpha^2)-g\sin(\theta-\alpha)\\
			-\frac{1}{2m}\rho VS_w(C_{L_{\alpha=0}}+C_{L_{\alpha}}\alpha+C_{L_{q}}\d\frac{c_A}{2V}q)+q+\frac{g\cos(\theta-\alpha)}{V}\\
			q\\
			\d\frac{1}{2I_y}\rho V^2S_w c_A (C_{M_{\alpha=0}}+C_{M_{\alpha}}\alpha+C_{M_{q}}\d\frac{c_A}{2V}q)\\
			V\sin(\theta-\alpha)\\
		\end{bmatrix}
	\end{split}
\end{equation*}
and
\begin{equation}\label{equ:C38}
	\small\begin{split}
		\bm{g}_n\big(\bm{x}_n\big)=\begin{bmatrix}
			0&\d\frac{1}{m}T_{\delta_t}\cos\alpha\\
			-\d\frac{1}{2m}\rho V S_wC_{L_{\delta_e}}&-\d\frac{1}{mV}T_{\delta_t}\sin\alpha\\
			0&0\\
			\d\frac{1}{2I_y}\rho V^2 S_wc_AC_{M_{\delta_e}}&0\\
			0&0\\
		\end{bmatrix}.
	\end{split}
\end{equation}
The vector fields {\small$\bm f_n: \mathbb{R}^5\times \mathbb{R}\longrightarrow\mathbb{R}^5 $}  and  {\small$\bm g_n: \mathbb{R}^5\longrightarrow\mathbb{R}^{5\times 2}$} are {\small$\mathcal{C}^k$} mapping with {\small$k\geq 0$}, meaning that they are  continuous vector-valued and matrix-valued functions of {\small$\bm x_n$}, and {\small$\bm{g}_n(\bm{x}_n)\neq\bm 0,~\forall \bm x_n$}.
\vspace*{-4mm}
\subsection{Control Objective}
Let us consider that the equilibrium point is denoted as {\small$(\bm x_e,\bm u_e,\xi_e)$} under a specific cruising flight condition for the system \eqref{equ:C5}, which satisfy 
{\small$
		\bm{f}_n(\bm{x}_e,\xi_e)+\bm{g}_n(\bm{x}_e)\bm{u}_e=\bm 0.
	$}
Define {\small$\bm x=\bm {x}_n-\bm x_e$}, {\small$\bm u=\bm {u}_n-\bm u_e$}, {\small$\xi=\xi_n- \xi_e$} and  it yields
\begin{equation}\label{equ:C6}
	\small\begin{split}
		\dot{\bm{x}}=&\dot{\bm {x}}_n=\bm{f}_n(\bm {x}_n,\xi_n)+\bm{g}_n(\bm {x}_n)\bm {u}_n\\
		=&\bm{f}_n(\bm {x}+\bm{x}_e,\xi+\xi_e)+\bm{g}_n(\bm {x}+\bm{x}_e)\bm{u}_e+\bm{g}_n(\bm {x}+\bm{x}_e)\bm{u}.
	\end{split}
\end{equation}
Thus, the model described by \eqref{equ:C6} 
can be reformulated as\footnote{According to \eqref{equ:C6}, it can be inferred that the origin {\small$\bm x = \bm 0$} serves as an equilibrium point for the system \eqref{equ:C7}  when {\small$\bm u = \bm 0$} and {\small$\xi=0$}, such that {\small$\bm f(\bm 0) = \bm 0$} whereas {\small$\bm{g}(\bm{x})\neq \bm 0$, $\forall \bm x\in\mathbb{R}^5$}.  Given the bounded open set {\small$\Omega\subseteq \mathbb{R}^{5}$}
	with {\small$\bm 0 \in \Omega$}, it follows that {\small$\bm x \in \Omega$}.}
\vspace*{-2mm}
\begin{equation}\label{equ:C7}
	\small\begin{split}
		&\dot{\bm{x}}=\bm{f}(\bm{x},\xi)+\bm{g}(\bm{x})\bm{u}
	\end{split}
\end{equation}
where 
{\small$\bm{f}(\bm{x},\xi)=\bm{f}_n(\bm {x}+\bm{x}_e,\xi+\xi_e)+\bm{g}_n(\bm {x}+\bm{x}_e)\bm{u}_e
$}
and  
{\small$\bm{g}(\bm{x})=\bm{g}_n(\bm {x}+\bm{x}_e).$} 
 In Section III,  we will detail how to train the representation function $\bm{\Phi}$ using DAIML.
The control objective of this study is to achieve continuous coordinated operation of structural deformation and flight maneuvering by designing  reasonable coordinated control strategies, so that the aircraft can reach a state of equilibrium, i.e.,\footnote{In this paper, we establish a tracking system  \eqref{equ:C7} that utilizes the equilibrium point of the original system \eqref{equ:C5} as the reference point. Thus, the equilibrium points are calculated in advance through LS fitting in \cite{yin2015coordinated}.} 
\vspace*{-2pt}
\begin{equation}\label{equ:C37}\small\lim_{t\rightarrow\infty}\bm x(t)=\bm 0.\end{equation}
\vspace*{-10mm}
\subsection{Problem Statement}
In most previous works \cite{lee2019linear,JIANG20151640,YUE2013909,seigler2009analysis,shi2015morphing}, structural deformation was directly applied to the MA as an external command, where the deformation process  was only related to time and remained unaffected  by the flight state. The attitude and motion control of the aircraft were still performed by conventional control mechanisms (elevators, throttle). However, in practice, we hope to enhance flight performance by improving the aerodynamic characteristics of the aircraft through active structural deformation.  
In these cases, the control problem becomes interesting if  the MA can autonomously change the wingspan length during flight to achieve coordinated control of deformation and flight.
Therefore, the problem under consideration in this study is to address the challenge of achieving coordinated control between deformation and flight to accomplish the flight mission.

To develop a differential game model for describing the coordinated control of deformation and flight, it is essential to decompose the unknown model {\small$\bm{f}(\bm{x},\xi)$}  into a linear combination {\small$\bm{\Phi}(\bm{x})\bm{a}(\xi)$}.  How to determine the  representation function {\small$\bm{\Phi}(\bm{x})$} and  linear coefficient {\small$\bm{a}(\xi)$}? Once the linear coefficient {\small$\bm{a}(\xi)$} is retrieved, a classifier still needs to be trained to get the specific deformation commands. Furthermore, upon deriving the differential game model, it is necessary to solve the CSDREs to obtain state-dependent (SD) feedback control solutions. 
\vspace*{-4mm}
\section{OFFLINE LEARNING STAGE}
For any
analytic function {\small$\bm f(\bm{x},\xi)$}, it can be decomposed into a {\small$\xi$}-invariant component {\small$\bm{\Phi}(\bm{x})$} and a {\small$\xi$}-dependent
component {\small$\bm{a}(\xi)$}  with arbitrary precision, where {\small$\bm{\Phi}(\bm{x})$} and {\small$\bm{a}(\xi)$} are both polynomials \cite{o2022neural}, i.e.,

{\small$~~~~~~~~~~~~~~~~~~~~~~~~~{\bm{f}}(\bm{x},\xi)\thickapprox\bm{\Phi}(\bm{x}){\bm{a}}(\xi)
	$}\\
where {\small$\bm{\Phi}(\bm{x})\in \mathbb{R}^{5\times h}$}  serves as  a basis or representation function that is applicable across all wingspan morphing 
conditions, and {\small$\bm a(\xi)\in \mathbb{R}^{h}$} denotes a set of linear coefficients that are  adjusted  
for each specific morphing condition.  

It is well-established that DNN architectures possess the capability to approximate {\small$\mathcal{C}^{0}(\bm \Omega)$} functions. In this section, the unmodeled dynamics term {\small$\bm{f}(\bm{x},\xi)$} is approximated using a single DNN.
In this case, {\small$\bm a$} denotes the weights of  the output-layer, and {\small$\bm \Phi$}
represents the hidden states preceding the output-layer, that is
\begin{equation}\label{equ:C36}
	\small\begin{split}
		{\bm{f}}(\bm{x},\xi)\approx\bm{W}^{T}\bm\varrho\big(\bm{\phi}(\bm{x})\big)&= \big(\bm I_5\otimes \bm\varrho^{T}\big({\bm{\phi}}(\bm{x})\big)\big){\rm vec}({\bm W})\\&=\bm{\Phi}(\bm{x}){\bm{a}}(\xi)
	\end{split}
\end{equation}
where {\small$\bm{W}=[\bm{w}_1~\bm{w}_2~\cdots~ \bm{w}_5]\in{\mathbb{R}^{m\times 5}}$} is the weight of its last layer,  {\small$\bm{\phi}(\bm{x}):\mathbb{R}^{5}\rightarrow{\mathbb{R}^p}$}
represents the 
inner-layer features of the DNN, and {\small$\bm\varrho:\mathbb{R}^{p}\rightarrow\mathbb{R}^{m} $} denotes an vector of activation functions for the last layer.  
Specifically,  {\small$\bm{\phi}(\bm{x})$} can be articulated as {\small$\bm{\phi}(\bm{x})=\bm \Xi_\iota\bm\varrho_\iota(\bm \Xi_{\iota-1}\bm\varrho_{\iota-1}(\bm \Xi_{\iota-2}\bm \varrho_{\iota-2}(\cdots\bm x)))$}, where {\small$\iota \in N$} indicates the number of inner-layers, {\small$\bm \Xi_\iota$} and {\small$\bm \varrho_\iota$} represent the corresponding  weights and activation functions, respectively.

In this part, the DAIML  is extended to learn a morphing condition independent DNN representation of the unmodeled dynamics term.  Different from conventional ML approaches, 
a discriminative network is  employed in training  to guarantee  that {\small$\bm \Phi (\bm x)$} is independent of variable-span  and that the   information about variable-span  is only contained
in {\small$\bm a (\xi)$} that is designed in the online game stage. 
The offline learning stage consists of two primary components: ML stage and classifier training stage. 
For the offline ML stage, the DAIML   is extended to learn a  morphing condition independent  representation  function {\small$\bm{\Phi}(\bm{x})$}. 
For the classifier training phase, a classifier is trained to represent the mapping relationship between states and morphing ratio.
\vspace*{-5mm}
\subsection{Data Collection}
To generate training data for learning the representation of unknown dynamics, the aircraft flies with the controller continuously excited by random noise for 3 minutes each in several different fixed wingspan conditions.
The set of input-output pairs corresponding to the $k$th trajectory is designated as the $k$th subdataset, denoted as {\small$D_{k }$}, with the specific wingspan morphing ratio {\small$\xi_k$}. 
Our dataset comprises six different subdatasets, with different wingspan morphing ratios from 0 to 1. The selected points are shown in Table I.
\begin{table}[h!]
	\vspace*{-2mm}
\begin{center}
		\caption{Points selection for MA.}
		\begin{tabular}{c|c|c|c|c|c|c} 
			\hline
			$\xi_n$ & 0.0& 0.2&0.4&0.6&0.8&1.0\\
			\hline
			b/m & 10.18&12.22&14.25&16.29&18.32&20.36 \\
			\hline
		\end{tabular}
	\end{center}
\end{table}
The data {\small$\bm x$} and {\small$\bm u$} are collected along each trajectory. Combining {\small$\bm g(\bm x)$} with \eqref{equ:C7}, a noisy measurement of the unknown dynamics, represented as  {\small$\bm y=\bm{f}+\bm d$}, is obtained, where {\small$\bm d$} encompasses all sources of noise. Hence,  we define  the dataset  {\small$\small\mathcal{D} = \{D_{1 }
,…, D_{6 }\}$}, where

	{\small$~~~~~~~~~~ D_{k }=\{\bm{x}_k(i),\bm{y}_{k}(i)=\bm{f}(\bm{x}_k(i),\xi_k)+\bm d\}^{N_k}_{i=1}$}\\
is {\small$N_k$} noisy input-output pairs with wingspan condition {\small$\xi_k$},  {\small$k$} represents the index for the morphing conditions, and  {\small$i$} denotes the index for the input-output pair.
\begin{assumption}
	There exist constants {\small$\overline{\bm a}$} and   {\small$\overline{\bm d}\in \mathbb{R}^{+}$} such that     {\small$||\bm a||\leq\overline{\bm a}$} and   {\small$||\bm d||\leq\overline{\bm d}$}.
\end{assumption}
\vspace*{-6mm}
\subsection{ML Algorithm}
The objective of ML is to learn a morphing condition independent representation {\small$\bm{\Phi}(\bm{x})$}, such that for any wingspan morphing ratio {\small$\xi$}, there exists coefficient 
{\small$\bm{a}(\xi)$} that enables the expression  {\small$\bm{\Phi}(\bm{x})\bm{a}(\xi)$} to effectively approximate the dynamics {\small$\bm{f}(\bm{x},\xi)$}. As shown in formula \eqref{equ:C36},  in this article, the representation function {\small$\bm{\Phi}(\bm{x})$} is regarded as   the inner-layer DNN features  with parameters {\small$\bm{\Xi}$}, where {\small$\bm{\Xi}$} represents  inner-layer weights of the DNN. Linear coefficient {\small$\bm a(\xi)$} is the output-layer weight of the vectorization.
Consequently, the problem can be framed as the following minimization problem:
\vspace*{-2mm}
\begin{equation}\label{equ:A11}
	\small\begin{split}
		\bm{\Xi}^{*}=\arg\min_{\bm{\Xi},\bm a_k}^{}\sum_{k=1}^{6}\sum_{i=1}^{N_k}(||\bm{y}_k(i)-\bm{\Phi}(\bm{x}_k(i),\bm{\Xi})\bm{a}_k||^2).
	\end{split}
\end{equation}

The training subdatasets {\small$D_{k}(k=1,\cdots,6)$}  are randomly extracted from the dataset  {\small$\mathcal{D}$}, where each {\small$D_k$} corresponds to a fixed morphing condition. We randomly sample two disjoint batches {\small$D_{k}^{\bm a}$}
and {\small$D_{k}^{\bm \Xi}$}  from  {\small$D_{k}$} with {\small$K$} and {\small$B$} data points, respectively, where {\small$D_{k}^{\bm a}\cap D_{k}^{\bm \Xi}=\emptyset$} and {\small$D_{k}^{\bm a}\cup D_{k}^{\bm \Xi}=D_{k}$}. Subsequently, for each dataset {\small$ D_{k}^{\bm a}$} with {\small$K$} data points (sampled from the same morphing condition $k$), the LS method is used to estimate $\bm a$ under each specific condition, i.e., adaptation
\begin{equation*}
	\small\begin{split}	
		&\underbrace{\begin{bmatrix}
				\bm{\Phi}(\bm{x}(1),\bm \Xi)\\
				\bm{\Phi}(\bm{x}(2),\bm \Xi)\\
				\vdots\\
				\bm{\Phi}(\bm{x}(K),\bm \Xi)\\
		\end{bmatrix}}\bm{a}=\underbrace{\begin{bmatrix}
				\bm{y}(1)\\
				\bm{y}(2)\\
				\vdots\\
				\bm{y}(K)\\
		\end{bmatrix}}.\\
		&~~\bm{\Phi}\in\mathbb{R}^{5K\times h}~~~~~~~\bm{y}\in\mathbb{R}^{5K\times1}
	\end{split}
\end{equation*}
Thus, the LS solution is given by
{\small$
	\bm{a}^{*}=(\bm{\Phi}^{T}\bm{\Phi})^{-1}\bm{\Phi}^{T}\bm{y}.
$}
With {\small$\bm a^{*}$} as an implicit function of {\small$\bm \Xi$},  another {\small$B$} data points  are used for  stochastic gradient descent (SGD) on {\small$\bm \Xi$} with loss 

	\noindent{\small$$~~
		\min_{\bm{\Xi}}^{}J(\bm a^{*} , \bm{\Xi},D_{k}^{\bm \Xi} )=\min_{\bm{\Xi}}^{}\sum_{i=1}^{B}\big(||\bm{y}_k(i)-\bm{\Phi}(\bm{x}_k(i),\bm{\Xi})\bm{a}^{*}||^2\big)
	$$}
i.e., 
{\small$
		\bm{\Xi}\leftarrow\bm{\Xi}-\beta\nabla_{\bm{\Xi}}J(\bm a^{*} , \bm{\Xi},D_{k}^{\bm \Xi} ),
	$}
where {\small$\beta\geq0$}  is a step size hyper-parameter.

The meta-objective
in \eqref{equ:A11} has the potential to result in overfitting and may not effectively get the morphing condition independent representation {\small$\bm{\Phi}(\bm x)$}. To address the issue of domain shift \cite{o2022neural}, the following 
adversarial optimization 
framework is given by 
\vspace*{-1.5mm}
\begin{equation}\label{equ:C39}
	\small\begin{split}
		&\max_{\bm h}^{}~~\min_{\bm{\Phi},\bm a}^{}\sum_{k=1}^{6}\sum_{i=1}^{N_k}\Big\{||\bm{y}_k(i)-\bm{\Phi}(\bm{x}_k(i),\bm{\Xi})\bm{a}_k||^2\\&~~~~~~~~~~-\alpha {\rm loss}
		\Big(\bm h\big(\bm{\Phi}(\bm{x}_k(i)\big),k\Big)\Big\}
	\end{split}
\end{equation}
where {\small$\bm h\in \mathbb{R}^{6}$} represents an additional  DNN that serves as a discriminator to predict the 
morphing condition index from {\small$\bm \Phi(\bm x)$}, {\small$\alpha\geq0$}  functions as  a hyper-parameter to regulate the degree of regularization. The output of DNN {\small$\bm h$} is a six-dimensional vector representing probability distribution for six fixed deformation.
loss$(\cdot)$  refers to  the cross entropy loss, i.e.,

\noindent{\small$$
		{\rm loss}\Big(\bm h\big(\bm{\Phi}(\bm{x}_k(i))\big),k\Big)=-\sum^{6}_{j=1}\delta_{kj}{\rm log}\big(\bm h^{T}\bm e_j\big)
	$$}where {\small$\delta_{kj}=1$}  if {\small$k=j$} and {\small$\delta_{kj}=0$} otherwise, and  {\small$\bm e_j$} is the standard basis function (e.g., {\small$\bm e_1=[1,0,0,0,0,0]^{T}$} ). The Algorithm 1 can be used to implement DAIML. \footnote{Intuitively, the DNNs 
			{\small$\bm h$} and {\small$\bm \Phi$} form a zero-sum max-min game. The DNN   {\small$\bm h$} aims to predict 
			the morphing condition index {\small$k$} directly from {\small$\bm \Phi$}, which is accomplished through the outer maximization. Conversely, the DNN {\small$\bm \Phi$} aims to approximate the label {\small$\bm{y}_k$}
			while complicating the task for {\small$\bm h$},  which is accomplished through the inner minimization. To put it another way, {\small$\bm h$} serves as  a learned regularizer 
			to eliminate  the variable-span information contained within {\small$\bm \Phi$}. }
\footnote{The normalization process (line 7) is implemented to ensure  {\small$||\bm a^{*}||\leq\gamma$}, 
			which improves the robustness of subsequent control design. The spectral normalization (line 9)   is employed to regulate the Lipschitz property of the DNN for improving generalization to new data \cite{shi2019neural,bartlett2017spectrally}. }
\footnote{The training process involves updating  {\small$\bm h$} and {\small$\bm{\Phi}(\bm{x})$} alternatively. Initially, {\small$\bm{\Phi}(\bm{x})$} is updated (line 9) while keeping {\small$\bm h$} fixed, followed by updating  {\small$\bm h$} (line 11) while keeping {\small$\bm{\Phi}(\bm{x})$} fixed.  The frequency parameter {\small$\eta$}, where {\small$0 < \eta \leq  1$},  is utilized to regulate the frequency of updates to the discriminator {\small$\bm h$}.}
		\vspace{-8pt}
\begin{algorithm}
\caption{The algorithm of DAIML  for MA}\label{alg:cap}
			\begin{algorithmic}[1]
				\Require
				{\small$\mathcal{D} = \{D_{\xi_1 },…, D_{\xi_6 }\}$}, {\small$\alpha\geq0$},~ {\small$0<\eta\leq1$},~{\small$\gamma>0$}
				\Ensure
				{\small$\bm{\Phi}$} and {\small$\bm h$};
				\State initialize {\small$\bm{\Phi}$} and {\small$\bm h$} randomly;
				\State repeat lines 3–12 until convergence
				\State extract {\small$D_{k }$} from {\small$\mathcal{D}$} randomly ;

				\State extract two disjoint batches {\small$D^{\bm a}_k$} and{\small $D^{\bm \Xi}_k$} from {\small$D_{k }$} randomly;
				\State solve the LS problem 
				{\small$\bm a^{*}=\arg\min_{\bm a}^{}\sum_{i\in D^{\bm a}_k}(||\bm{y}_k(i)-\bm{\Phi}(\bm{x}_k(i),\bm{\Xi})\bm{a}||^2)$};
				\If {$\|\bm a^{*}\|>\gamma$}
				\State {\small$\d\bm a^{*}=\gamma\frac{\bm a^{*}}{\|\bm a^{*}\|}$}
				\EndIf
				\State train DNN {\small$\bm{\Phi}$}  with loss {\small$\sum_{i\in D^{\bm \Xi}_k}\Big\{||\bm{y}_k(i)-\bm{\Phi}\big(\bm{x}_k(i),\bm{\Xi}\big)\bm{a}^{*}||^2-\alpha {\rm loss}
				\Big(\bm h\big(\bm{\Phi}(\bm{x}_k(i))\big),k\Big)\Big\}$} by SGD and spectral normalization
				\If {$rand()<\eta$} 
				\State train DNN {\small$\bm{h}$} with loss {\small$\sum_{i\in D^{\bm \Xi}_k}{\rm loss}
				\Big(\bm h\big(\bm{\Phi}(\bm{x}_k(i))\big),k\Big)$} by SGD 
				\EndIf

			\end{algorithmic}
\end{algorithm}
	\vspace{-1pt}
		\vspace*{-6mm}
		\subsection{Training of Task Classifier}
		To provide specific morphing commands to the aircraft, the linear coefficient {\small$\bm a$} must be converted into the wingspan morphing ratio {\small$\xi$}. Hence, the next goal is to obtain a mapping relationship between {\small$\bm \chi_k(i)=[\bm f_k^{T}(i)~\bm x_k^{T}(i)]^{T}$} and the wingspan morphing  ratio {\small$\xi$}.
		
		Here, a DNN with weights {\small$\bm{\Theta}$} will be trained as a task classifier with {\small$\bm\chi \in \mathbb{R}^{10}$} as input and {\small$\bm\rho\in \mathbb{R}^6$} as output. The output {\small$\bm\rho=[\rho_1,~\rho_2,~\rho_3,~ \rho_4,~\rho_5, ~\rho_6]$} refers to the probability distribution for each fixed deformation.  The loss function is represented by the following cross-entropy: 
		\vspace{-2pt}
\begin{equation*}
			\small\begin{split}
				{\rm loss}\Big(\bm \rho\big(\bm \chi_k(i)\big),k\Big)=
				-\sum^{6}_{j=1}\delta_{kj}{\rm log}\Big(\bm \rho^{T}\bm e_j\Big).
			\end{split}
		\end{equation*}
		
The following estimation of wingspan morphing ratio  is ultimately derived  by multiplying the probability distribution under fixed deformation with the six fixed wingspan morphing ratios:
		{\small$
				\hat{\xi}=\sum^{6}_{k=1}\xi_k  \rho_k.
			$}
		\vspace*{-4mm}
		\section{ONLINE GAME PHASE}

In this part, a non-cooperative game  framework is introduced to design the parameter {\small$\bm a$}. Through non-cooperative game theory, the aircraft model and flight control can form coordinated control with each other to realize the autonomous morphing control of the MA.  
		
	\vspace*{-4mm}	
\subsection{Differential Game Model for Describing the Coordinated
			Control of Deformation and Flight}
		
		Based on learned representation function {\small$\bm \Phi (\bm x)$}, the unknown dynamics term {\small$\bm{f}(\bm{x},\xi)$}  is approximated by the following linear
		combination:
		\vspace*{-2mm}	
\begin{equation}\label{equ:C11}
			\small\begin{split}
				\bm{f}(\bm{x},\xi)\approx\bm{\Phi}(\bm{x}){\bm a(\xi)}
			\end{split}
		\end{equation}	
where {\small$\bm a(\xi)$} is treated as a morphing control operated by {\small$\xi$} and will be updated during the game phase.
		The substitution of \eqref{equ:C11} into \eqref{equ:C7} yields
	
			{\small$
				~~~~~~~~~~~~~~~~~~~~~~~\dot{\bm{x}}=\bm{\Phi}(\bm{x})\bm{a}+\bm g(\bm{x})\bm{u}.$}\\
		Furthermore, the following equivalent model can be derived:
		\vspace*{-2mm}
\begin{equation}\label{equ:C12}
			\small\begin{split}
				&\dot{\bm{x}}=\bm{A}(\bm{x})\bm{x}+\bm{B}_{\bm{a}}(\bm{x})\bm{a}+\bm{B}_{\bm{u}}(\bm{x})\bm{u},~~\bm x(0) =\bm x_0
			\end{split}
\end{equation}
		where {\small$\bm{A}(\bm{x})\bm{x}=\bm 0$}, {\small$\bm{B}_{\bm{a}}(\bm{x})=\bm{\Phi}(\bm{x})$} and {\small$\bm{B}_{\bm{u}}(\bm{x})=\bm g(\bm{x})$}.
		
		The coordinated control of MA is treated as a non-cooperative game between morphing control and flight control,  both of which are regarded as rational agents.  The cost functions associated with the {\small$\bm{a}$} 
		and {\small$\bm{u}$}  are provided as follows:
		\vspace*{-2mm}
\begin{equation}\label{equ:C14}
			\small\begin{split}
				&J_{\bm{u}}\big(\bm x_0,\bm{u},\bm{a}\big)=\frac{1}{2}\int_{0}^{\infty}\{||\bm{x}(t)||^{2}_{\bm Q_{\bm{u}}(\bm{x})} +||\bm{u}(t)||^{2}_{\bm R_{\bm{u}}(\bm{x})} \}dt 
			\end{split}
		\end{equation}
\vspace*{-4mm}
		\begin{equation}\label{equ:A10}
			\small\begin{split}
				&J_{\bm{a}}\big(\bm x_0,\bm{u},\bm{a}\big)=\frac{1}{2}\int_{0}^{\infty}\{ ||\bm{x}(t)||^{2}_{\bm Q_{\bm{a}}(\bm{x})}+||\bm{a}(t)||^{2}_{\bm R_{\bm{a}}(\bm{x})}\}dt.
			\end{split}
		\end{equation}
The weighting matrices are {\small$\mathcal{C}^{0}(\bm \Omega)$} matrix-valued functions satisfying
		{\small$\bm{Q}_{\bm{u}}(\bm{x})=\bm{Q}^{T}_{\bm{u}}(\bm{x})\geq\bm 0$}, {\small$\bm{Q}_{\bm{a}}(\bm{x})=\bm{Q}^{T}_{\bm{a}}(\bm{x})\geq\bm 0$}, {\small$\bm{R}_{\bm{u}}(\bm{x})(\bm{x})=\bm{R}^{T}_{\bm{u}}(\bm{x})>\bm 0$} and {\small$\bm{R}_{\bm{a}}(\bm{x})=\bm{R}^{T}_{\bm{a}}(\bm{x})>\bm 0,~\forall \bm x\in \Omega$}. 
		\begin{assumption}
			{\small$\bm{A}(\bm{x})$}, {\small$\bm{B}_{\bm{a}}(\bm{x})$} and {\small$\bm{B}_{\bm{u}}(\bm{x})$} are  {\small$\mathcal{C}^{0}$} matrix-valued functions.  The matrix pairs {\small$\big(\bm{A}(\bm{x}),\bm{B}_{\bm l}(\bm{x}),\sqrt{\bm Q_{\bm l}(\bm{x})},\bm l=\bm{u},\bm{a}\big)$} are pointwise stabilizable and 
			detectable.
		\end{assumption}
		
		The optimal solution to the above problem results in the so-called Nash 
		optimal strategies {\small$\bm{u}^{*}$} and {\small$\bm{a}^{*}$} satisfying 
\vspace*{-5mm}
\begin{equation*}
			\small\begin{split}
				J_{\bm{u}}\big(\bm x_0,\bm{u}^{*},\bm{a}^{*}\big)\leq J_{\bm{u}}\big(\bm x_0,\bm{u},\bm{a}^{*}\big),~
				J_{\bm{a}}\big(\bm x_0,\bm{u}^{*},\bm{a}^{*}\big)\leq J_{\bm{a}}\big(\bm x_0,\bm{u}^{*},\bm{a}\big).
			\end{split}
		\end{equation*}
		For the  linearlike structure in  $\eqref{equ:C12}$, by mimicking the linear quadratic differential game theory formulation, we get the following 
		the closed-loop Nash optimal 
		strategies:  
\vspace*{-1.5mm}
		\begin{equation}\label{equ:C13}
			\small\begin{split}
				&\bm{u}=-\bm{R}_{\bm{u}}^{-1}(\bm{x})\bm{B}^{T}_{\bm{u}}(\bm{x})\bm P_{\bm{u}}(\bm x)\bm{x}\\&\bm{a}=-\bm{R}_{\bm{a}}^{-1}(\bm{x})\bm{B}^{T}_{\bm{a}}(\bm{x})\bm P_{\bm{a}}(\bm x)\bm{x}
			\end{split}
		\end{equation}
		where {\small$\big(\bm P_{\bm{u}}(\bm x)$, $\bm P_{\bm{a}}(\bm x)\big)$} are unique, positive semidefinite for each {\small$\bm x \in \Omega$} and  correspond to the pointwise stabilizing solutions of the following CSDREs:
	\vspace*{-1.5mm}	
\begin{equation}\label{equ:A1}
			\small\begin{split}
				&{\rm sym}\Big(\bm P_{\bm u}(\bm x) \bm{A}_{\bm{a}}\big(\bm P_{\bm{a}}(\bm{x})\big)\Big)+\bm\Gamma_{\bm u}(\bm P_{\bm  u}(\bm x))
=\bm 0
			\end{split}
		\end{equation}
		\vspace*{-4.35mm}
\begin{equation}\label{equ:A2}
			\small\begin{split}
				&{\rm sym}\Big(\bm P_{\bm a}(\bm x) \bm{A}_{\bm{u}}\big(\bm P_{\bm{u}}(\bm{x})\big)\Big)+ \bm\Gamma_{\bm a}(\bm P_{\bm  a}(\bm x))
=\bm 0
			\end{split}
		\end{equation}
		where {\small$\bm{A}_{\bm{u}}\big(\bm P_{\bm{u}}(\bm{x})\big)=\bm{A}(\bm{x})-\bm{B}_{\bm{u}}(\bm x)\bm{R}_{\bm{u}}^{-1}(\bm{x})\bm{B}^{T}_{\bm{u}}(\bm{x})\bm P_{\bm{u}}(\bm x)$} and {\small$ \bm{A}_{\bm{a}}\big(\bm P_{\bm{a}}(\bm x)\big)=\bm{A}(\bm{x})-\bm{B}_{\bm{a}}(\bm{x})\bm{R}_{\bm{a}}^{-1}(\bm{x})\bm{B}^{T}_{\bm{a}}(\bm{x})\bm P_{\bm{a}}(\bm x)$}, {\small$\bm\Gamma_{\bm u}\big(\bm P_{\bm  u}(\bm x)\big)=-\bm P_{\bm{u}}(\bm x)\bm{B}_{\bm{u}}(\bm{x})\bm{R}^{-1}_{\bm{u}}(\bm x)\bm{B}_{\bm{u}}^{T}(\bm{x})\bm P_{\bm{u}}(\bm x)+\bm Q_{\bm{u}}(\bm x)$}, and {\small$\bm\Gamma_{\bm a}\big(\bm P_{\bm  a}(\bm x)\big)=-\bm P_{\bm{a}}(\bm x)\bm{B}_{\bm{a}}(\bm{x})\bm{R}^{-1}_{\bm{a}}(\bm x)\bm{B}_{\bm{a}}^{T}(\bm{x})\bm P_{\bm{a}}(\bm x)+\bm Q_{\bm{a}}(\bm x)$}.
		By substituting control laws  \eqref{equ:C13} into equation \eqref{equ:C12}, the following closed-loop dynamics is provided as
		\vspace*{-2mm}	
\begin{equation}\label{equ:C15}
			\small\begin{split}
				\dot{\bm{x}}(t)=\bm A_{c}\big(\bm P_{\bm{u}}(\bm x),\bm P_{\bm{a}}(\bm x)\big)\bm x.
			\end{split}
		\end{equation}
		where {\small$\bm A_{c}\big(\bm P_{\bm{u}}(\bm x),\bm P_{\bm{a}}(\bm x)\big)=\bm{A}(\bm{x})-\bm{B}_{\bm{u}}(\bm x)\bm{R}_{\bm{u}}^{-1}(\bm{x})\bm{B}^{T}_{\bm{u}}(\bm{x})\bm P_{\bm{u}}(\bm x)-\bm{B}_{\bm{a}}(\bm{x})\bm{R}_{\bm{a}}^{-1}(\bm{x})\bm{B}^{T}_{\bm{a}}(\bm{x})\bm P_{\bm{a}}(\bm x)$.} Under Assumption 2, the CSDRE control laws are pointwise stabilizing, such that  the closed-loop  matrix {\small$\bm A_{c}\big(\bm P_{\bm{u}}(\bm x),\bm P_{\bm{a}}(\bm x)\big)$} given
		in \eqref{equ:C15} is pointwise Hurwitz {\small$\forall \bm x$}.\footnote{Consider the linearlike structure \eqref{equ:C12} with 
			feedback \eqref{equ:C13}, and suppose that Assumption 2 holds. Then, the origin of the system
			\eqref{equ:C15} is asymptotically stable in  {\small$\Omega \in \mathbb{R}^{5}$}. If {\small$\|e^{\bm A_{c}\big(\bm P_{\bm{u}}(\bm x),\bm P_{\bm{a}}(\bm x)\big)t}\|\leq M$} for some real {\small$M>0$} and {\small$\forall \bm x \in \Omega \in \mathbb{R}^{5}$}, {\small$\forall t \in \mathbb{R}^{+}$}, then the system  \eqref{equ:C15} is globally asymptotically stable in {\small$\Omega$} \cite{2012Survey}.}	 When necessary, the arguments  {\small$t$} and {\small$\bm x$} are dropped for notational simplicity, for example, {\small$\bm A(\bm x)=\bm A$}.	
		\vspace*{-4mm}
		\subsection{Lyapunov Iterations for Solving CSDREs}
		Building upon the work of  \cite{Li1995LyapunovIF}, an algorithm to calculate a
		pair of stabilizing solutions {\small$(\bm P_{\bm{u}},\bm P_{\bm a})$}   of the
		CSDREs  is proposed. 
		Given Assumption 2 and assumption  {\small$\bm a=\bm0$},  a 
		unique positive semidefinite solution of auxiliary SDRE
		\vspace*{-5mm}
\begin{equation}\label{equ:A3}
			\small\begin{split}
				&{\rm sym}\big(\bm P_{\bm u}^{(0)} \bm{A}\big)+\bm\Gamma_{\bm u}(\bm P_{\bm  u}^{(0)})=\bm 0.
			\end{split}
		\end{equation}
		exists such that {\small$\bm{A}_{\bm{u}}\big(\bm P_{\bm{u}}^{(0)}\big)$} is pointwise Hurwitz. By substituting {\small$\bm P_{\bm{u}}=\bm P_{\bm{u}}^{(0)}$} 
		into \eqref{equ:A2}, we get the second SDRE  as 
\vspace*{-4mm}
\begin{equation}\label{equ:A4}
			\small\begin{split}
				{\rm sym}\big(\bm P_{\bm a}^{(0)} \bm{A}_{\bm{u}}(\bm P^{(0)}_{\bm{u}})\big)+ \bm\Gamma_{\bm a}(\bm P_{\bm  a}^{(0)})
=\bm 0
			\end{split}
		\end{equation}
		Since {\small$\bm{A}_{\bm{u}}(\bm P_{\bm{u}}^{(0)})$} is pointwise Hurwitz and {\small$\bm Q_{\bm{a}}=\bm Q_{\bm{a}}^{T}\geq \bm 0$},  there exists a unique, positive semidefinite {\small$\bm P_{\bm{a}}^{(0)}$} renders the corresponding closed-loop matrix  {\small$\bm A_{c}(\bm P_{\bm{u}}^{(0)},\bm P_{\bm{a}}^{(0)})$}  pointwise Hurwitz, {\small$\forall \bm x$}.  Subsequently, 
		{\small$\bm P_{\bm{u}}^{(0)}$} and {\small$\bm P_{\bm{a}}^{(0)}$},  are used to initialize our algorithm. The following algorithm is provided for solving the CSDREs.\\
		
\vspace*{-2mm}
		\noindent\textbf{Algorithm 2 }\\
		\noindent $\bm{Step~1}$: Start with initial matrix {\small$(\bm P_{\bm a}^{(0)},\bm P_{\bm u}^{(0)})$} such 
		{\small$
				\bm A_{c}(\bm P_{\bm{u}}^{(0)},\bm P_{\bm{a}}^{(0)})
			$}
		is pointwise Hurwitz. Let {\small$i=0$}.
		\vspace*{1mm}

		\noindent $\bm{Step~2}$: For {\small$i\geq 0$}, solve the following  SD Lyapunov equations for {\small$\bm P_{\bm{u}}^{(i+1)}$} and {\small$\bm P_{\bm{a}}^{(i+1)}$}
		\vspace*{-2.1mm}
\begin{equation}\label{equ:A5}
			\small\begin{split}
				{\rm sym}\big(\bm P_{\bm{u}}^{(i+1)}\bm A_{c}(\bm P_{\bm{u}}^{(i)},\bm P_{\bm{a}}^{(i)})\big)+\bm Y_{\bm u}(\bm P_{\bm{u}}^{(i)})= \bm 0
			\end{split}
		\end{equation}
	\vspace*{-4mm}	
\begin{equation}\label{equ:C33}
			\small\begin{split}
				{\rm sym}\big(\bm P_{\bm{a}}^{(i+1)}\bm A_{c}(\bm P_{\bm{u}}^{(i)},\bm P_{\bm{a}}^{(i)})\big)+\bm Y_{\bm a}(\bm P_{\bm{a}}^{(i)})= \bm 0 
			\end{split}
		\end{equation}
	with initial conditions {\small$\bm P_{\bm a}^{(0)}$} and {\small$\bm P_{\bm u}^{(0)}$},  where	{\small$\bm Y_{\bm u}(\bm P_{\bm{u}}^{(i)})=\bm P_{\bm{u}}^{(i)}\bm{B}_{\bm{u}}\bm{R}^{-1}_{\bm{u}}\bm{B}_{\bm{u}}^{T}\bm P_{\bm{u}}^{(i)}+\bm Q_{\bm{u}},$} {\small$\bm Y_{\bm a}(\bm P_{\bm{a}}^{(i)})=\bm P_{\bm{a}}^{(i)}\bm{B}_{\bm{a}}\bm{R}^{-1}_{\bm{a}}\bm{B}_{\bm{a}}^{T}\bm P_{\bm{a}}^{(i)}+\bm Q_{\bm{a}}$ }
		
		\vspace*{1mm}
		\noindent $\bm{Step~3}$: Let
		{\small$
				\bm u^{(i+1)}=-\bm{R}_{\bm{u}}^{-1}\bm{B}^{T}_{\bm{u}}\bm P_{\bm{u}}^{(i+1)}\bm x,
			$}
		{\small$
				\bm a^{(i+1)}=-\bm{R}_{\bm{a}}^{-1}\bm{B}^{T}_{\bm{a}}\bm P_{\bm{a}}^{(i+1)}\bm x 
			$}
		and repeat Step 2 with {\small$i+1$} .
		
		Before discussing the convergence of the Algorithm 2, it is essential to introduce the following lemma, 
		which states that the control laws {\small$\bm u^{(i+1)}$}  and {\small$\bm a^{(i+1)}$} are improvements over 
		{\small$\bm u^{(i)}$}  or {\small$\bm a^{(i)}$}, respectively. 
		\begin{lemma}
			Let 
			{\small$
					\bm{u}^{(i)}=-\bm{R}_{\bm{u}}^{-1}\bm{B}^{T}_{\bm{u}}\bm P_{\bm{u}}^{(i)}\bm x$},
					{\small$\bm{a}^{(i)}=-\bm{R}_{\bm{a}}^{-1}\bm{B}^{T}_{\bm{a}}\bm P_{\bm{a}}^{(i)}\bm x$}
			be such that {\small$\bm A_{c}(\bm P_{\bm{u}}^{(i)},\bm P_{\bm{a}}^{(i)})$} is pointwise Hurwitz, {\small$\forall \bm x$}.
			For an arbitrary {\small$\mathcal{T}>0$}, we define
			\begin{equation*}
				\small\begin{split}
				\bm{u}^{(\mathcal{T})}=\left\{\begin{aligned}
					-\bm{R}_{\bm{u}}^{-1}\big(\bm{x}(t)\big)\bm{B}^{T}_{\bm{u}}\big(\bm{x}(t)\big)&\bm P_{\bm{u}}^{(i+1)}\big(\bm x(t)\big)\bm x(t),~~t\in [0,\mathcal{T}]\\
					-\bm{R}_{\bm{u}}^{-1}(\bm{x}(t))\bm{B}^{T}_{\bm{u}}\big(\bm{x}(t)\big)&\bm P_{\bm{u}}^{(i)}\big(\bm x(t)\big)\bm x(t),~~t\in (\mathcal{T},\infty)
				\end{aligned}\right.\end{split}
			\end{equation*}
			where {\small$\bm P_{\bm{u}}^{(i+1)}(\bm x)$} is given by \eqref{equ:A5}. Then  
			\begin{equation*}
				\small\begin{split}
					J_{\bm{u}}\big(\bm x_0,\bm{u}^{(\mathcal{T})},\bm{a}^{(i)}\big)&=\frac{1}{2}\int_{0}^{\infty}\{ ||\bm{x}||^{2}_{\bm Q_{\bm{u}}}+||\bm{u}^{(\mathcal{T})}||^{2}_{\bm R_{\bm{u}}} \}dt\\& \leq J_{\bm{u}}\big(\bm x_0,\bm{u}^{(i)},\bm{a}^{(i)}\big).
				\end{split}
			\end{equation*}
			In particular
			\begin{equation*}
				\small\begin{split}
					&J_{\bm{u}}(\bm x_0,\bm{u}^{(\mathcal{T})},\bm{a}^{(i)})=
					J_{\bm{u}}(\bm x_0,\bm{u}^{(i)},\bm{a}^{(i)})
					\\& -\frac{1}{2}\int_{0}^{\mathcal{T}} \bm{x}^{T}(\bm P_{\bm{u}}^{(i+1)}-\bm P_{\bm{u}}^{(i)})\bm{B}_{\bm{u}}\bm{R}_{\bm{u}}^{-1}\bm{B}^{T}_{\bm{u}}(\bm P_{\bm{u}}^{(i+1)}-\bm P_{\bm{u}}^{(i)})\bm{x}dt
				\end{split}
			\end{equation*}
			where 
			{\small$
					\dot{\bm{x}}(t)=\bm{A}_{\bm{a}}(\bm P_{\bm{a}}^{(i)})\bm x+\bm{B}_{\bm{u}}\bm{u}^{(\mathcal{T})}.
				$}
{\small$\bm{a}^{(\mathcal{T})}$} and {\small$J_{\bm{a}}(\bm x_0,\bm{u}^{(i)},\bm{a}^{(\mathcal{T})})$} are similar to it.	
		\end{lemma}

		\noindent\emph{Proof~of~Lemma~1.} Note that\footnote{
			Based on Algorithm 2, we define
			
{\small$~~~~~~~~V_{\bm u}^{(i)}=J_{\bm{u}}\big(\bm{x},\bm{u}^{(i)},\bm{a}^{(i)}\big),~V_{\bm a}^{(i)}=J_{\bm{a}}\big(\bm{x},\bm{u}^{(i)},\bm{a}^{(i)}\big).
                $}\\
                \vspace*{-2mm}
			{\small($\d\frac{\partial  V_{\bm{l}}^{(i)}}{\partial \bm x}$}, {\small$\bm l=\bm a,\bm u$)} are  written in SD coefficient forms \cite{2012Survey}
			\vspace*{-1mm}
\begin{equation}\label{equ:C22}
				\small\begin{split}		
					(\frac{\partial  V_{\bm{u}}^{(i)}}{\partial \bm x})^{T}=\bm P_{\bm{u}}^{(i+1)}\bm x
				\end{split}
			\end{equation}
\vspace*{-2mm}
			\begin{equation}\label{equ:C23}
				\small\begin{split}
					(\frac{\partial  V_{\bm{a}}^{(i)}}{\partial \bm x})^{T}=\bm P_{\bm{a}}^{(i+1)}\bm x	
				\end{split}
			\vspace*{-5mm}
\end{equation}
			for some matrix-valued functions {\small$\bm P_{\bm{a}}^{(i+1)}$} and {\small$\bm P_{\bm{u}}^{(i+1)}$}. }
		\footnote{
			If  the equations \eqref{equ:C22} and \eqref{equ:C23} hold,  then {\small$(V_{\bm a}^{(i)}(\bm x)$, $V_{\bm u}^{(i)}(\bm x))$} have the following standard forms
			(see Lemma 2-22 in \cite{Berger1968PerspectivesIN})
			\vspace*{-1.5mm}
\begin{equation*}
				\small\begin{split}
					&V_{\bm u}^{(i)}(\bm x)=\bm x^{T}\int_{0}^{1}\bm P_{\bm{u}}^{(i+1)}(\bm x\sigma)\sigma d\sigma\bm x\\
					&V_{\bm a}^{(i)}(\bm x)=\bm x^{T}\int_{0}^{1}\bm P_{\bm{a}}^{(i+1)}(\bm x\sigma)\sigma d\sigma\bm x.
				\end{split}
			\end{equation*}}
		\begin{equation}\label{equ:C27}
			\small\begin{split}
				J_{\bm{u}}\big(\bm x_0,&\bm{u}^{(\mathcal{T})},\bm{a}^{(i)}\big)=
				\frac{1}{2}\int_{0}^{\infty}\{ ||\bm{x}||^{2}_{\bm Q_{\bm{u}}}+||\bm{u}^{(\mathcal{T})}||^{2}_{\bm R_{\bm{u}}}\}dt
\d\\&-\int_{0}^{\mathcal{T}} \frac{dV_{\bm{u}}^{(i)}}{dt} dt
				+\int_{0}^{\mathcal{T}}\frac{\partial  V_{\bm{u}}^{(i)}}{\partial \bm x}\frac{d   \bm x }{  dt}dt.\\
				&=\frac{1}{2}\int_{0}^{\infty}\{ ||\bm{x}||^{2}_{\bm Q_{\bm{u}}}+||\bm{u}^{(\mathcal{T})}||^{2}_{\bm R_{\bm{u}}}\}dt
				-\\&\int_{0}^{\mathcal{T}} \frac{d}{dt}\Big(\bm x^{T}\int_{0}^{1}\bm P_{\bm{u}}^{(i+1)}(\bm x\sigma)\sigma d\sigma\bm x\Big)dt
				+\\&\frac{1}{2}\int_{0}^{\mathcal{T}}\Big\{ 2\bm{x}^{T}\bm P_{\bm{u}}^{(i+1)}\bm A_{c}(\bm P_{\bm{u}}^{(i+1)},\bm P_{\bm{a}}^{(i)})\bm x\Big\}dt
			\end{split}
		\end{equation}
		The second item in \eqref{equ:C27} equals
		\begin{equation*}
			\small\begin{split}
				\bm{x}^{T}(\mathcal{T})&\int_{0}^{1}\bm P_{\bm{u}}^{(i+1)}\big(\bm x(\mathcal{T})\sigma\big)\sigma d\sigma\bm{x}(\mathcal{T})\\&-\bm{x}^{T}_0\int_{0}^{1}\bm P_{\bm{u}}^{(i+1)}\big(\bm x_0\sigma\big)\sigma d\sigma\bm x_0
			\end{split}
		\end{equation*}
		while the first item in \eqref{equ:C27} equals 
		\begin{equation*}
			\small\begin{split}
				&\frac{1}{2}\int_{0}^{\mathcal{T}}\bm{x}^{T}\bm Y_{\bm u}(\bm P_{\bm{u}}^{(i+1)})\bm xdt+\bm{x}^{T}(\mathcal{T})\int_{0}^{1}\bm P_{\bm{u}}^{(i+1)}\big(\bm x(\mathcal{T})\sigma\big)\sigma d\sigma\bm{x}(\mathcal{T}).
			\end{split}
		\end{equation*}
		Thus,
		\vspace*{-2mm}
{\small\begin{align}\label{equ:C34}
			\begin{split}
				&J_{\bm{u}}(\bm x_0,\bm{u}^{(\mathcal{T})},\bm{a}^{(i)})=
				\bm{x}^{T}_0\int_{0}^{1}\bm P_{\bm{u}}^{(i+1)}\big(\bm x_0\sigma\big)\sigma d\sigma\bm x_0
				+\\&\d\frac{1}{2}\int_{0}^{\mathcal{T}} \Big\{ \bm{x}^{T}\Big(\bm Q_{\bm{u}}+\bm P_{\bm{u}}^{(i+1)}\bm{B}_{\bm{u}}\bm{R}_{\bm{u}}^{-1}\bm{B}_{\bm{u}}^{T}\bm P_{\bm{u}}^{(i+1)}\\&+{\rm sym}\big( \bm P_{\bm{u}}^{(i+1)}\bm A_{c}(\bm P_{\bm{u}}^{(i+1)},\bm P_{\bm{a}}^{(i)})\big)
				\Big)\bm x\Big\}dt
			\end{split}
		\end{align}}
		From \eqref{equ:A5}, we have
		\begin{equation*}
			\small\begin{split}
				\bm Q_{\bm{u}}=-{\rm sym}\big( \bm P_{\bm{u}}^{(i+1)}\bm A_{c}(\bm P_{\bm{u}}^{(i)},\bm P_{\bm{a}}^{(i)})\big)
				-\bm P_{\bm{u}}^{(i)}\bm{B}_{\bm{u}}\bm{R}_{\bm{u}}^{-1}\bm{B}_{\bm{u}}^{T}\bm P_{\bm{u}}^{(i)}
			\end{split}
		\end{equation*}
		and, after some  algebraic manipulations, the second integrand in \eqref{equ:C34} becomes 
		\begin{equation*}
			\small\begin{split}
				-\bm{x}^{T}(\bm P_{\bm{u}}^{(i+1)}-\bm P_{\bm{u}}^{(i)})\bm{B}_{\bm{u}}\bm{R}_{\bm{u}}^{-1}\bm{B}^{T}_{\bm{u}}(\bm P_{\bm{u}}^{(i+1)}-\bm P_{\bm{u}}^{(i)})\bm{x}.
			\end{split}
		\end{equation*}
		Similarly, for morphing controller $\bm a$, we can get a similar result.   $\qedsymbol$
		
		Next, we show the convergence of  Algorithm 2.
		\begin{theorem}
			Let {\small$\bm P_{\bm{u}}$}  and {\small$\bm P_{\bm{a}}$} are  the stabilizing solutions of CSDREs \eqref{equ:A1}-\eqref{equ:A2}. {\small$\{\bm P_{\bm{u}}^{(i)}\}$}  and {\small$\{\bm P_{\bm{a}}^{(i)}\}$} are the sequences generated by Algorithm 2. Then, when {\small$i\rightarrow \infty$},  {\small$\bm P_{\bm{u}}^{(i)}\rightarrow \bm P_{\bm{u}}$} and {\small$\bm P_{\bm{a}}^{(i)}\rightarrow \bm P_{\bm{a}}$}. 
		\end{theorem}
		
		
		\noindent\emph{Proof~of~Theorem~1.} 
		For the optimization problem defined by \eqref{equ:C12} and \eqref{equ:C14}-\eqref{equ:A10}, corresponding Hamiltonians for each control agent are defined as
		\begin{equation}\label{equ:C16}
			\small\begin{split}
				H_{\bm u}&(\bm x,\bm u, \bm a^{*}, \frac{\partial  V_{\bm u}^{*}}{\partial \bm x})=\frac{1}{2}\big(||\bm{x}||^{2}_{\bm Q_{\bm{u}}}+||\bm{u}||^{2}_{\bm R_{\bm{u}}}\big)\\&+\frac{\partial  V_{\bm u}^{*}}{\partial \bm x}\{\bm{A}\bm{x}+\bm{B}_{\bm{a}}\bm{a}^{*}+\bm{B}_{\bm{u}}\bm{u}\}
			\end{split}
		\end{equation}
	\vspace*{-1.8mm}
\begin{equation}\label{equ:C17}
			\small\begin{split}
				H_{\bm a}&(\bm x,\bm u^{*}, \bm a, \frac{\partial  V^{*}_{\bm a}}{\partial \bm x})=\frac{1}{2}\big(||\bm{x}||^{2}_{\bm Q_{\bm{a}}}+||\bm{a}||^{2}_{\bm R_{\bm{a}}}\big)\\&+\frac{\partial  V^{*}_{\bm a}}{\partial \bm x}\{\bm{A}\bm{x}+\bm{B}_{\bm{a}}\bm{a}+\bm{B}_{\bm{u}}\bm{u}^{*}\}
			\end{split}
		\end{equation}
		where {\small$V_{\bm u}^{*}$} and {\small$V_{\bm a}^{*}$} are   value functions and defined by
		\begin{equation}\label{equ:C18}
			\small\begin{split}
				V_{\bm u}^{*}=\inf\limits_{\bm{u}}J_{\bm{u}}(\bm{x},\bm{u}(\cdot),\bm{a}^{*}(\cdot))
			\end{split}
		\end{equation}
\vspace*{-6mm}
		\begin{equation}\label{equ:C19}
			\small\begin{split}
				V_{\bm a}^{*}=\inf\limits_{\bm{a}}J_{\bm{a}}(\bm{x},\bm{u}^{*}(\cdot),\bm{a}(\cdot)).
			\end{split}
		\end{equation}
		The necessary conditions for the Nash optimal strategies are given as
		\begin{equation*}
			\small\begin{split}
				\bm u^{*}&=\arg \min_{\bm u}\{H_{\bm u}(\bm x,\bm u, \bm a^{*},\frac{\partial  V^{*}_{\bm{u}}}{\partial \bm x})\}
				=-\bm{R}_{\bm{u}}^{-1}\bm{B}^{T}_{\bm{u}}(\frac{\partial  V^{*}_{\bm{u}}}{\partial \bm x})^{T}
			\end{split}
		\end{equation*}
		\begin{equation*}
			\small\begin{split}
				\bm a^{*}&=\arg \min_{\bm a}^{}\{H_{\bm a}(\bm x,\bm u^{*}, \bm a,\frac{\partial  V^{*}_{\bm{a}}}{\partial \bm x})\}
				=-\bm{R}_{\bm{a}}^{-1}\bm{B}^{T}_{\bm{a}}(\frac{\partial  V^{*}_{\bm{a}}}{\partial \bm x})^{T}.
			\end{split}
		\end{equation*}
		The method of successive approximations in dynamic programming, as applied to equations \eqref{equ:C12} and \eqref{equ:C16}-\eqref{equ:C19}, consists of two steps.

		(i) Take any initial control laws {\small$\bm a^{(0)}$} and {\small$\bm u^{(0)}$} from \eqref{equ:A3} and \eqref{equ:A4}, that is,
		{\small
				$\bm u^{(0)}=-\bm{R}_{\bm{u}}^{-1}\bm{B}^{T}_{\bm{u}}\bm P_{\bm{u}}^{(0)}\bm x
			$} and
		{\small
				$\bm a^{(0)}=-\bm{R}_{\bm{a}}^{-1}\bm{B}^{T}_{\bm{a}}\bm P_{\bm{a}}^{(0)}\bm x,
			$}
		with {\small$\bm P_{\bm{u}}^{(0)} $} and {\small$\bm P_{\bm{a}}^{(0)} $} being pointwise positive semidefinite, and the corresponding performance criterion is
		\begin{equation}\label{equ:C20}
			\small\begin{split}
				V_{\bm{u}}^{(0)}(\bm x)=&\frac{1}{2}\int_{t}^{\infty}\Big\{ ||\bm{x}(\tau)||_{\bm Q_{\bm{u}}\big(\bm{x}(\tau)\big)}^{2}+||\bm{\bm{u}}^{(0)}\big(\bm{x}(\tau)\big)||_{\bm{R}_{\bm{u}}\big(\bm{x}(\tau)\big)}^{2}\Big\}d\tau\\=&
				\frac{1}{2}\int_{t}^{\infty} \bm{x}^{T}(\tau)\bm Y_{\bm u}\Big(\bm P_{\bm{u}}^{(0)}\big(\bm x(\tau)\big)\Big)\bm{x}(\tau) d\tau
			\end{split}
		\end{equation}
		along the trajectories of the system 
		\begin{equation}\label{equ:C29}
			\small\begin{split}
				\dot{\bm{x}}=&\bm{A}\bm{x}+\bm{B}_{\bm{a}}\bm a^{(0)}+\bm{B}_{\bm{u}}\bm u^{(0)}=\bm A_{c}\big(\bm P_{\bm{u}}^{(0)},\bm P_{\bm{a}}^{(0)}\big)\bm x.
			\end{split}
		\end{equation}
		Similarly, the performance criterion  for {\small$\bm a^{(0)}$} is
		\begin{equation}\label{equ:C21}
			\small\begin{split}
				V_{\bm{a}}^{(0)}(\bm x)=\frac{1}{2}\int_{t}^{\infty} \bm{x}^{T}(\tau)\bm Y_{\bm a}\Big(\bm P_{\bm{a}}^{(0)}\big(\bm x(\tau)\big)\Big)\bm{x}(\tau)  d\tau.
			\end{split}
		\end{equation}
		From \eqref{equ:C20} and \eqref{equ:C21}, we can derive the expressions for {\small$\d\frac{\partial  V^{(0)}_{\bm u}}{\partial \bm x}$} and {\small$\d\frac{\partial  V^{(0)}_{\bm a}}{\partial \bm x}$} along with equation \eqref{equ:C29}. The determination of these expressions will be addressed at a later stage. 
		
		(ii)According to \eqref{equ:C16}, the Hamiltonian {\small$\d H_{\bm u}(\bm x,\bm u, \bm a^{*}, \frac{\partial  V_{\bm u}^{(0)}}{\partial \bm x})$} with {\small$\d\frac{\partial  V^{(0)}_{\bm u}}{\partial \bm x}$ } is provided 
		and a new approximation for $\bm u$ can be obtained by minimizing the Hamiltonian with respect to  $\bm u$, i.e.,
		\begin{equation}\label{equ:C25}
			\small\begin{split}
				\bm u^{(1)}=-\bm{R}_{\bm{u}}^{-1}\bm{B}^{T}_{\bm{u}}(\frac{\partial  V_{\bm u}^{(0)}}{\partial \bm x})^{T}.
			\end{split}
		\end{equation}
		Similarly, the stabilizing control for the {\small$\bm a$} as
		{\small$
				\d\bm a^{(1)}=-\bm{R}_{\bm{a}}^{-1}\bm{B}^{T}_{\bm{a}}(\frac{\partial  V_{\bm a}^{(0)}}{\partial \bm x})^{T}.
			$}
		Then, {\small$\d\frac{\partial V^{(0)}_{\bm{u}}}{\partial \bm x}$} can be calculated from \eqref{equ:C20} and  \eqref{equ:C29}. That is,
		under  {\small$\bm u^{(0)}$} and {\small$\bm a^{(0)}$}, we have
		\begin{equation*}
			\small\begin{split}
				\frac{dV_{\bm{u}}^{(0)}}{dt}=\frac{\partial V_{\bm{u}}^{(0)}}{\partial \bm x}\frac{d\bm x}{dt}=&\frac{\partial  V_{\bm u}^{(0)}}{\partial \bm x}\bm A_{c}\big(\bm P_{\bm{u}}^{(0)},\bm P_{\bm{a}}^{(0)}\big)\bm x\\=&-\frac{1}{2}\bm{x}^{T}\bm Y_{\bm u}\big(\bm P_{\bm{u}}^{(0)}\big)\bm x.
			\end{split}
		\end{equation*}
		From \eqref{equ:C22}, {\small$\d\frac{\partial V^{(0)}_{\bm{u}}}{\partial \bm x}$} can be written in the following form: 
		\begin{equation}\label{equ:C26}
			\small\begin{split}
				\frac{\partial  V_{\bm{u}}^{(0)}}{\partial \bm x}=\bm x^{T}\bm P_{\bm{u}}^{(1)}
			\end{split}
		\end{equation}
		and, we get

		{\small$~~~~~~~~~\d-\frac{1}{2}\bm{x}^{T}\bm Y_{\bm u}(\bm P_{\bm{u}}^{(0)})\bm x=\bm x^{T}\bm P_{\bm{u}}^{(1)}\bm A_{c}(\bm P_{\bm{u}}^{(0)},\bm P_{\bm{a}}^{(0)})\bm x.
			$}
		\\Using the standard symmetrization technique, we obtain
		\begin{equation}\label{equ:C24}
		\small\begin{split}
				&{\rm sym} \big(\bm P_{\bm{u}}^{(1)}\bm A_{c}(\bm P_{\bm{u}}^{(0)},\bm P_{\bm{a}}^{(0)})\big)=-\bm Y_{\bm u}(\bm P_{\bm{u}}^{(0)}).
			\end{split}
		\end{equation}
		Due to the fact that {\small $\bm A_{c}(\bm P_{\bm{u}}^{(0)},\bm P_{\bm{a}}^{(0)})$} is  pointwise Hurwitz and {\small$-\bm Y_{\bm u}(\bm P_{\bm{u}}^{(0)})$} is pointwise negative semidefinite, it can  be concluded that a unique, pointwise positive semidefinite solution  {\small$\bm P_{\bm{u}}^{(1)}$} exists. Assuming the 
		design parameters {\small$\bm Q_{\bm{u}}>\bm 0$} , it follows that the corresponding solution of \eqref{equ:C24} 
		will be also positive definite. From \eqref{equ:C25} and \eqref{equ:C26}, we have  
		{\small$
				\bm u^{(1)}=-\bm{R}_{\bm{u}}^{-1}\bm{B}^{T}_{\bm{u}}\bm P_{\bm u}^{(1)}\bm x.
			$}
		Performing similar operations   for the morphing control {\small$\bm a^{(1)}$}, we 
		get
		{\small$
				\bm a^{(1)}=-\bm{R}_{\bm{a}}^{-1}\bm{B}^{T}_{\bm{a}}\bm P_{\bm a}^{(1)}\bm x.
			$}
		
		Repeating above two steps with {\small$\bm u^{(1)}$} and {\small$\bm a^{(1)}$}, we derive the subsequent values 
		{\small$\bm u^{(2)}$} and {\small$\bm a^{(2)}$} along with the corresponding matrices  {\small$\bm P_{\bm u}^{(2)}$} and {\small$\bm P_{\bm a}^{(2)}$}. By following the same 
		procedure, the sequences of the solution matrices can be derived. In addition, according  to Lemma 1, the stabilizing control laws {\small$\bm u^{(i+1)}$}  and {\small$\bm a^{(i+1)}$} are improvements over 
		{\small$\bm u^{(i)}$}  and {\small$\bm a^{(i)}$}, respectively, even when it is used for just a finite time interval, i.e.,
		
{\small$
				~~~~~~~~~~~~~~~~~~~\bm P_{\bm a}^{(i+1)}\leq\bm P_{\bm a}^{(i)},~
				\bm P_{\bm u}^{(i+1)}\leq\bm P_{\bm u}^{(i)}.
			$}\\
		Thus, {\small$\lim_{i\rightarrow\infty}^{}\bm P_{\bm l}^{(i)}=\bm P_{\bm l}^{(\infty)}(\bm l=\bm a,\bm u)$} exist  by a theorem on 
		monotonic convergence of nonnegative operators \cite{Kantorovich1952FunctionalAI}. By taking the limits of \eqref{equ:A5} and \eqref{equ:C33}  as {\small$i\rightarrow\infty$} we 
		obtain
		\vspace*{-1mm}
\begin{equation*}
			\small\begin{split}
				&{\rm sym} \big(\bm P_{\bm{u}}^{(\infty)} \bm A_{c}(\bm P_{\bm{u}}^{(\infty)},\bm P_{\bm{a}}^{(\infty)})\big)
				+\bm Y_{\bm u}\big(\bm P_{\bm{u}}^{(\infty)}\big) = \bm 0\\ 
				&{\rm sym} \big(\bm P_{\bm{a}}^{(\infty)} \bm A_{c}(\bm P_{\bm{u}}^{(\infty)},\bm P_{\bm{a}}^{(\infty)})\big)
				+\bm Y_{\bm a}\big(\bm P_{\bm{a}}^{(\infty)}\big) = \bm 0.
			\end{split}
\end{equation*}		
Since {\small$\bm P_{\bm u}$} and {\small$\bm P_{\bm a}$} are the unique, positive semidefinite 
		solutions of \eqref{equ:A1} and \eqref{equ:A2}, respectively, {\small$\bm P_{\bm{u}}^{(\infty)}=\bm P_{\bm u}$} and {\small$\bm P_{\bm{a}}^{(\infty)}=\bm P_{\bm a}$. $\qedsymbol$}
		\vspace*{-3mm}
		\subsection{Implementation of  Online Game Phase }
		The complexity of   the Lyapunov iterations   \eqref{equ:A5} and \eqref{equ:C33}   precludes the possibility of obtaining 
		analytical solutions except in certain simplistic low-dimensional systems. 
		Hence, to achieve real-time operation, the analytical solutions are replaced with the numerical solutions at each step. 
		That is, for
		any  specified value of  {\small$\bm x$}, the online algorithm involves simply calculating the
		positive semidefinite solutions   {\small$\bm P_{\bm{u}}$},   {\small$\bm P_{\bm{a}}$} to the algebraic CSDREs
		\eqref{equ:A1} and   \eqref{equ:A2} and applying the control \eqref{equ:C13} at that   {\small$\bm x$}. 
		This approach is much more appealing and computationally efficient than finding the analytical solutions.
		The specific steps are outlined in the following algorithm:\\
\vspace*{-3mm}
		
\noindent\textbf{Algorithm 3 }\\
		$\bm{Step~1}$: A common representation   {\small$\bm{\Phi}(\bm{x})$} and DNN   {\small$\bm \rho(\bm{\chi})$} are obtained through offline  learning phase. Given the design threshold   {\small$\varepsilon$} and the sampling period   {\small$T$}.
		Let   {\small$i=0$}.
\vspace*{1mm}
		
\noindent$\bm{Step~2}$:  {\small$\bm P_{\bm{a}}^{(0)}$} and {\small$\bm P_{\bm{u}}^{(0)}$} are chosen based on present moment:\\
		\noindent i) if {\small$t=0$}, {\small$\bm P_{\bm{a}}^{(0)}$} and {\small$\bm P_{\bm{u}}^{(0)}$} are obtained by solving \eqref{equ:A3}-\eqref{equ:A4} with {\small$\bm x_0$};\\
		\noindent ii) if {\small$t>0$}, and the previous step size {\small$\bm P_{\bm{a}}$} and {\small$\bm P_{\bm{u}}$} make corresponding closed loop matrix 
		{\small$
				\bm A_{c}\big(\bm P_{\bm{u}},\bm P_{\bm{a}}\big)
			$}
		stable in present step, then {\small$\bm P^{(0)}_{\bm{u}}$} and {\small$\bm P^{(0)}_{\bm{a}}$} are the previous step size {\small$\bm P_{\bm{u}}$} and {\small$\bm P_{\bm{a}}$}, otherwise solve  algebraic Riccati equations \eqref{equ:A3}-\eqref{equ:A4} at 
		the present moment to get {\small$\bm P^{(0)}_{\bm{u}}$} and {\small$\bm P^{(0)}_{\bm{a}}$}, and let
			{\small$
				\bm{u}^{(0)}=-\bm{R}_{\bm{u}}^{-1}\bm{B}^{T}_{\bm{u}}\bm P_{\bm{u}}^{(0)}\bm{x},~~\bm{a}^{(0)}=-\bm{R}_{\bm{a}}^{-1}\bm{B}^{T}_{\bm{a}}\bm P_{\bm{a}}^{(0)}\bm{x}
			$}.

		\vspace*{1mm}
		\noindent $\bm{Step~3}$: Solving Lyapunov iterations \eqref{equ:A5} and \eqref{equ:C33} with current state {\small$\bm x$}, we can  get {\small$\bm P^{(i+1)}_{\bm{u}}$} and {\small$\bm P^{(i+1)}_{\bm{a}}$}.
		
\vspace*{1mm}
		\noindent$\bm{Step~4}$: Turn to Step 6 if the following criterion is 
		satisfied
		
{\small$~~~~~~~~\max \{ \|\bm P_{\bm{u}}^{(i+1)}-\bm P_{\bm{u}}^{(i)}\|, \|\bm P_{\bm{a}}^{(i+1)}-\bm P_{\bm{a}}^{(i)}\|\}\leq \varepsilon,
			$}\\
		otherwise go to Step 5.
		\vspace*{1mm}

		\noindent $\bm{Step~5}$: If the time taken for all previous steps exceeds  the sample period {\small$T$}, turn to Step 6; otherwise, {\small$\bm P_{\bm{u}}^{(i)}=\bm P_{\bm{u}}^{(i+1)}$} and {\small$\bm P_{\bm{a}}^{(i)}=\bm P_{\bm{a}}^{(i+1)}$}  turn to Step 3.
		\vspace*{1mm}

		\noindent$\bm{Step~6}$: The calculation in  current sampling period is accomplished, let {\small$\bm P_{\bm u}=\bm P_{\bm u}^{(i+1)}$} and  {\small$\bm P_{\bm a}=\bm P_{\bm a}^{(i+1)}$} to derive the input vector

		{\small$
				\bm{u}^{(i+1)}=-\bm{R}_{\bm{u}}^{-1}\bm{B}^{T}_{\bm{u}}\bm P_{\bm{u}}^{(i+1)}\bm{x},~~\bm{a}^{(i+1)}=-\bm{R}_{\bm{a}}^{-1}\bm{B}^{T}_{\bm{a}}\bm P_{\bm{a}}^{(i+1)}\bm{x}
			$}\\
		and then, we can get  the real-time dynamic {\small$\bm \chi=[\bm x^{T}~\bm f^{T}]^{T}$} and the morphing ratio estimate is obtained by multiplying the probability distribution under fixed deformation with the fixed wingspan morphing  ratio, that is,
		{\small$\hat{\xi}=\sum^{6}_{k=1}\xi_{k}\rho_k.$}

\vspace*{1mm}
		\noindent $\bm{Step~7}$: Enter the next step size, turn to Step 2.
		\vspace*{-4mm}
		\section{SIMULATION STUDY}
		This section considers the MA model in \cite{yin2015coordinated}. The parameters of MA are reorganized in Table II and Table III. It is assumed that the  
control mechanism parameters  are operated in the domain {\small$U=\{\bm  u_n\in \mathbb{R}^2:-0.7~{ rad}\leq\delta_e\leq 0.7~{ rad},~0\%\leq\delta_t\leq100\%\}$}.		
\begin{table}[h!]
       \vspace*{-5mm} 
        \begin{center}
				\caption{Parameters of MA}
				\begin{tabular}{l c} 
					\hline
					Parameters &Value\\
					\hline
		          {Wing airfoil}                   &{NACA 643-618}\\
                    Weight of MA $ m({\rm kg})$& 1247\\
					Constant of gravity $ g({\rm m/s^2})$& 9.8\\
					Shortest wingspan $b_{\rm min}({\rm m})$ &10.18\\ 
					Longest wingspan $b_{\rm max}({\rm m})$ &20.36\\ 
					Reference area $S_{w}({\rm m^{2}})$& 17.09\\
					Mean aerodynamic chord $c_{A}({\rm m})$& 1.74\\
					 {Rolling moment of inertia $I_x({\rm kg.m^2})$}& {1420.9}\\
                    Pitch moment of inertia $I_y ({\rm kg.m^2})$&4067.5\\
                   { Yaw moment of inertia $I_z ({\rm kg.m^2})$}&{4786.0}\\ 
					{ Product of inertia $I_{xy},I_{yz},I_{zx} ({\rm kg.m^2})$}&{0}\\ 
                    Thrust coefficient $T_{\delta_t} ({\rm N/\%})$&21.3\\
					\hline
				\end{tabular}
			\end{center}
		\end{table}
\vspace*{0mm} 
{\small\begin{table}[h!]
			\vspace*{-5mm} 
\begin{center}
				{\small\caption{{Approximate expression of longitudinal aerodynamic parameters in \eqref{equ:C3} }}}
				{\begin{tabular}{l l} 
					\hline
					Coefficient &Approximate expression\\
					\hline
					$C_{L_{\alpha=0}}$                   &$0.0098Ma+0.4890\xi_n+0.3340$\\
                    $C_{L_{\alpha}}$& $-0.0001h-1.0597Ma+6.0872\xi_n+5.9792$\\
					$C_{L_{\delta_e}}$& $-0.0013h+0.0316Ma+0.4099$\\
					$C_{L_{q}}$ &$0.8710Ma+5.1386\xi_n+9.6995$\\ 
					$C_{D_{\alpha=0}}$ &$0.0005h-0.0277Ma+0.0142\xi_n+0.0288$\\ 
					$C_{D_{\alpha}}$& $0.0001h+0.0325Ma+0.0906\xi_n+0.1883$\\
					$C_{D_{\alpha^2}}$& $-0.0011h-1.2434Ma+0.1408\xi_n+2.1775$\\
					$C_{M_{\alpha=0}}$&$-0.0001h+0.0031Ma-0.2436\xi_n+0.0121$\\
                    $C_{M_{\alpha}}$&$-0.0001h-0.0922Ma-1.4954\xi_n-1.6444$\\
                    $C_{M_{\delta_e}}$&$0.0030h-0.1256Ma-0.9766$\\ 
					 $C_{M_{q}}$&$-0.6857Ma-1.0762\xi_n-18.1012$\\ 
					\hline
				\end{tabular}}
			\end{center}
		\vspace*{-2mm}
\end{table}	}

The process of  simulation is:  initially, the MA has a  height of {\small$4990~m$} and a  velocity of {\small$35~m/s$}. The initial
 value of attack angle, pitch angle, and pitch angular velocity are
 respectively set to be {\small$0.1968 ~rad$}, {\small$0.1729~rad$}, and {\small$0~rad/s$}, i.e.,
 {\small $\alpha = 0.1968~rad$}, {\small$\theta = 0.1729~rad$}, and {\small$q = 0~rad/s$}  at {\small$t = 0~s$}, i.e., {\small$\bm x_n(0)=[35~~0.1968~~0.1729~~0~~4990]^{T}$} (An aircraft in level flight at a constant altitude and speed generally has a positive  attack angle and a positive pitch angle, with a pitch  angular velocity of zero).
  Assuming {\small$\xi_e=0.2$},  the equilibrium points can be calculated in advance through LS fitting in \cite{yin2015coordinated}, namely
				{\small$\bm x_e= [40~~0.1268~~0.1259~~0~~5000]^{T}$}
and				{\small$\bm u_e=[-0.2890~~42.8188]^{T}$}.
		By designing the   coordinated control of deformation and flight, the aircraft can ascend to an altitude of {\small$5000~m$}  and reach a flight velocity of {\small$40~m/s$}  rapidly. And then, the aircraft   enters a new level flight phase, maintaining stable flight velocity and attitude throughout the process, thus achieving the control objective \eqref{equ:C37}. This section studies the effectiveness of the coordinated control strategies designed for MA in completing trajectory tracking through numerical simulation experiments. It verifies the rationality and superiority of the proposed control strategies and analyzes the improvement of the coordinated control strategies on the maneuverability of the aircraft. In order to better simulate real-world scenarios, we add noise into the simulation  \cite{LinPayload}, specifically the noise corresponds to a Gaussian 
distribution with a variance of {\small$30\%$} of the actuator. 
		\vspace*{-5mm}
		\subsection{Data Collection and Network Training}
		
{In this paper, MATLAB is used for numerical simulation of the flight process of a MA to collect training data. We allow a MA to fly for {\small$5$} minutes along random trajectories under six different morphing conditions, with the wingspan morphing ratio ranging from {\small$0$} to {\small$1$}. Data are sampled at {\small $100~Hz$}, generating a total of {\small$30000$} data points.
 We collect data {\small $\bm x$} and {\small$\bm u$} along each trajectory, and use numerical differentiation to compute {\small$\dot{\bm x}$}.   Combining {\small$\bm g(\bm x)$} with \eqref{equ:C7}, the measurement {\small $\bm y$} can be obtained.}

		The DNN in \eqref{equ:C36} has five fully connected hidden layers, with an architecture {\small$5\rightarrow  64 \rightarrow64 \rightarrow32 \rightarrow 5\rightarrow 5$} and  the last layer activation function {\small$\bm\varrho$} is linear transfer function. The  inner-layer DNN features  {\small$\bm \phi (\bm x)$} have an architecture {\small$5\rightarrow  64 \rightarrow64 \rightarrow32 \rightarrow 5$} and  tangent sigmoid activation.  The five components of the unknown dynamic, {\small$f_1$}, {\small$f_2$}, {\small$f_3$}, {\small$f_4$}
		and {\small$f_5$}, exhibit a high degree of correlation and possess shared characteristics,  so {\small$\bm \phi$} is  employed as the basis function for all the components.  Hence, the 
		unknown dynamic {\small$\bm f$} is approximated by
		\vspace{-5pt}
\begin{equation*}
			{\small\begin{split}
				&\bm{f}(\bm{x},\xi)\approx\bm \Phi(\bm x)\bm a=\big(\bm I_5\otimes \bm\varrho^{T}\big({\bm{\phi}}(\bm{x})\big)\big){\rm vec}({\bm W})=\\
				&\begin{bmatrix}
					\bm{\phi}^{T}(\bm{x})&0 &0&0&0\\
					0&\bm{\phi}^{T}(\bm{x})& 0&0&0\\
					0&0& \bm{\phi}^{T}(\bm{x})&0&0\\
					0&0&0& \bm{\phi}^{T}(\bm{x})&0\\
					0&0&0&0& \bm{\phi}^{T}(\bm{x})\\
\end{bmatrix}\begin{bmatrix}
					\bm{w}_1\\
					\bm{w}_2\\
					\bm{w}_3\\
					\bm{w}_4\\
					\bm{w}_5\\
				\end{bmatrix}.
			\end{split}}
		\vspace{-5.6pt}
\end{equation*}
		
The DNN {\small$\bm h $} has  an architecture {\small$5\rightarrow  128  \rightarrow 6$} and tangent sigmoid activation. The classifier DNN {\small$\bm \rho $} has  an architecture {\small$10\rightarrow  200  \rightarrow 200 \rightarrow 128  \rightarrow 6$} and tangent sigmoid activation. The appropriate hyper-parameters  are  selected, i.e., the discriminator training frequency {\small$\eta$} is given as {\small$0.5$},  the normalization constant {\small$\gamma$} is given as {\small$10$} and the  degree of regularization {\small$\alpha$} is given as {\small$0.1$}.\footnote{The frequency $0<\eta<1$  is utilized to regulate the frequency of updates to the discriminator $\bm h$. It is important to note that when $\eta = 1$, both the generator and the discriminator are updated during each iteration. For the purpose of enhancing training stability, we adopt $\eta=0.5$, a value that is frequently employed in the training of generative adversarial networks  \cite{good2014}. The regularization parameter $\alpha >0$. Note that $\alpha=0$ corresponds to the non-adversarial ML case which does not incorporate the adversarial regularization term in \eqref{equ:C39}. Clearly, a 
proper choice of $\alpha$ can effectively avoid over-fitting. However, if $\alpha$ is too high, it may degrade prediction performance. Therefore, we recommend 
using relatively small value for $\alpha$ such as 0.1.} 
		We follow Algorithm 1 to train {\small$\bm \Phi$}. The predicted {\small$\hat{\bm f}$}
		is compared to the measured {\small${\bm y}$} for another part of the
		validation trajectory. The verification results of flight altitude and velocity are shown in  Figs. 3. The results presented in Figs. 3 indicate that once  {\small$\bm  a$} have adapted to the current
		morphing condition, the NN is {capable} of effectively fitting the observed data. 
\begin{figure*}[h]
			\centering
			\includegraphics[width=17.5cm, height=6.5cm]{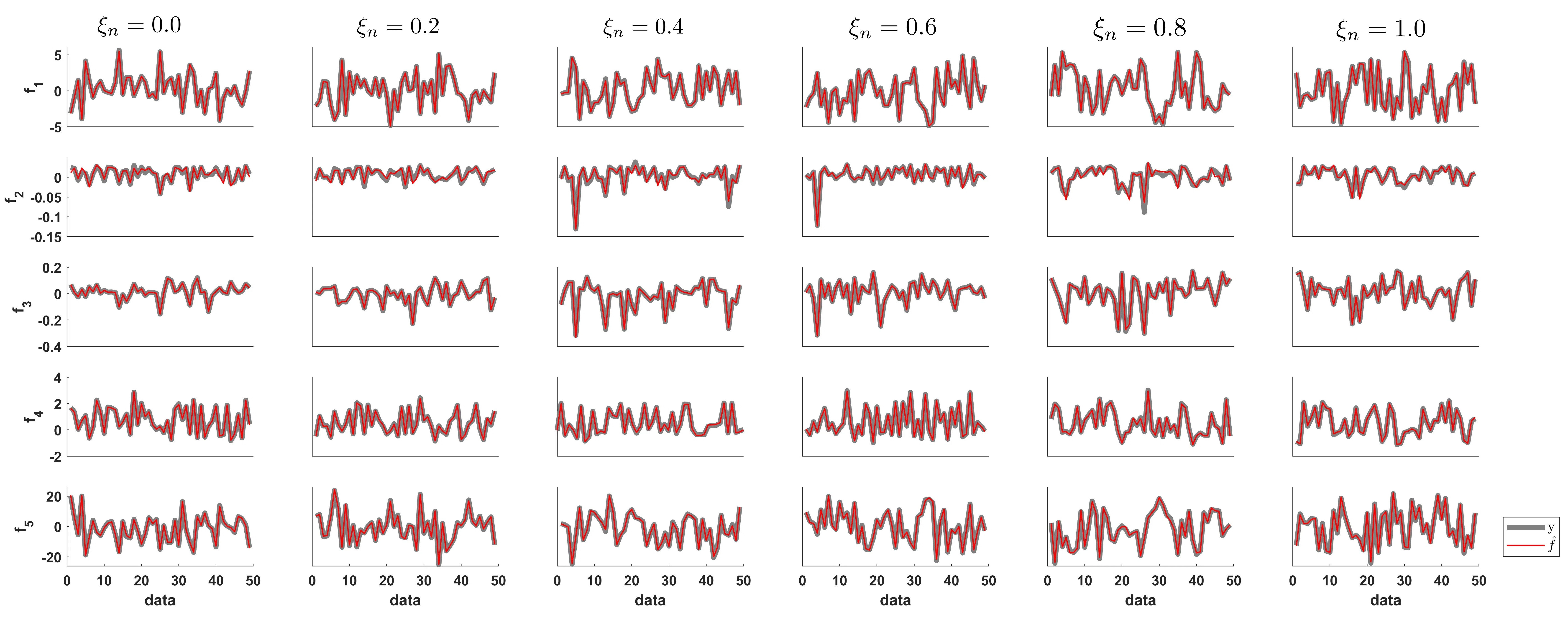}
			\caption{{\small$\mathbf{Measured~ unknown ~ model~versus~dynamic~estimation~based~on~DAIML.}$ This figure shows the comparison between model estimation $\hat {\bm f}$ and measured values $\bm y$ under different deformation conditions. $f_1,...,f_5$ respectively represent the vector elements of unmodeled dynamics $\bm f$. The results indicate that once  $\bm  a$ have adapted to the current morphing condition, the NN is capable of effectively fitting the observed data.}}
			\label{fig8}
\vspace{-15pt}
		\end{figure*}
\vspace{-6mm}
\subsection{Coordinated Control of Deformation and Flight}
		\vspace{-2mm}
The weighting functions in cost functions \eqref{equ:C14} and \eqref{equ:A10} are selected as 
		{\small$ \bm Q_{\bm u}$} $=$ {\small$\bm Q_{\bm a}$} {\small$={\rm diag}\{20,2000, 700,200,2\}$}, {\small$\bm R_{\bm u}$} {\small$={\rm diag}\{1500,0.25\}$},~{\small$\bm R_{\bm a}$} {\small$={\rm diag}\{100,100,100,100,100,$} {\small$100,100,100,100,100,100,100,100,100,100,200,200,$}{\small$100,1,3\}.$}\footnote{For weights {\small$\bm Q_{\bm u}$}, {\small$\bm Q_{\bm a}$}, {\small $\bm R_{\bm u}$} and  {\small$\bm R_{\bm a}$}, a higher weight is allocated to the state variables of flight velocity, attack angle, pitch angle, and pitch angular velocity, to ensure precise control of these important state variables and avoid flight instability. In this context, the expression $\dot{h}=V\sin(\theta-\alpha)$ means that the update of flight altitude depends on flight velocity, attack angle, and pitch angle. Hence, a smaller weight is assigned to flight altitude. A higher weight is assigned to the $\delta_e$ to suppress excessive control surface manipulation and ensure a relatively stable altitude control, while a  relatively smaller weight value for $\delta_t$ and wing deformation parameter $\bm a$ is selected to achieve rapid response and maximum the merit of wingspan deformation.}
	The Algorithm 3
		is used to calculate the
		real-time {\small$\bm P_{\bm u}^{(i)}$} and {\small$\bm P_{\bm a}^{(i)}$}.
		The control inputs {\small$\bm u_n$} and morphing command {\small$\xi_n$} are shown in Fig. 4.  The states {\small$\bm x_n$} are shown in Fig. 5. One can observe that the trajectories of state  converge to equilibrium
points, which means 
		control objectives \eqref{equ:C37} is achieved by the designed coordinated control strategies. The above  simulation results demonstrate that the designed control laws effectively stabilize  the states to the  equilibrium points.  The dynamic process is good and the values of the control inputs are within a reasonable range.
		
		\vspace*{-5mm}
		\begin{figure}[h]
			\centering
			\includegraphics[width=8.5cm, height=5.8cm]{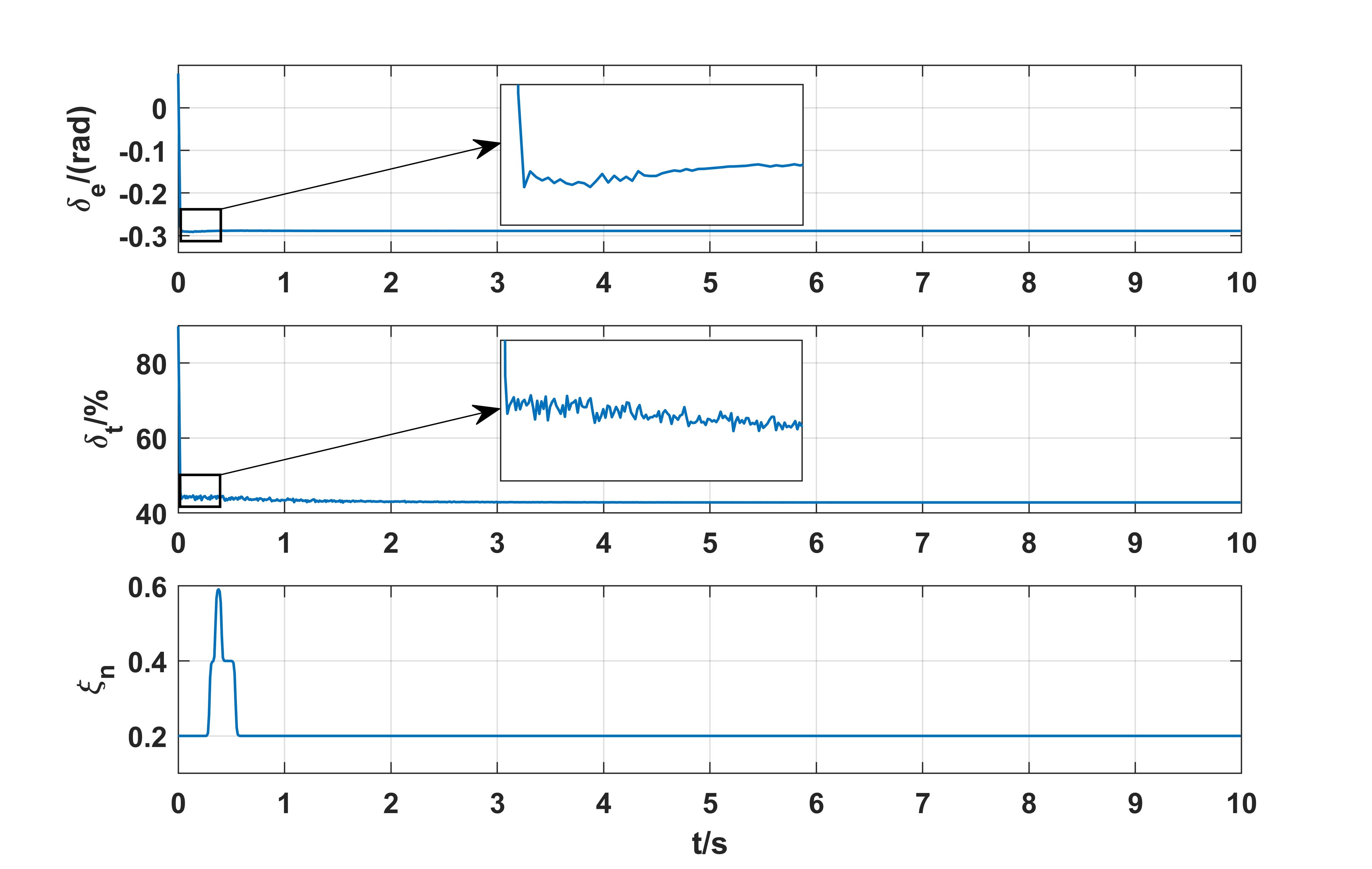}
			\caption{{\small$\mathbf{Control}$ $\mathbf{inputs}$ $\mathbf{\bm u_n}$ $\mathbf{and}$ $\mathbf{morphing}$ $\mathbf{command}$ $\mathbf{\xi_n.}$ }}
			\label{fig8}
\vspace{-20pt}
		\end{figure}
		\begin{figure}[h]
			\centering
			\includegraphics[width=8.2cm, height=8.5cm]{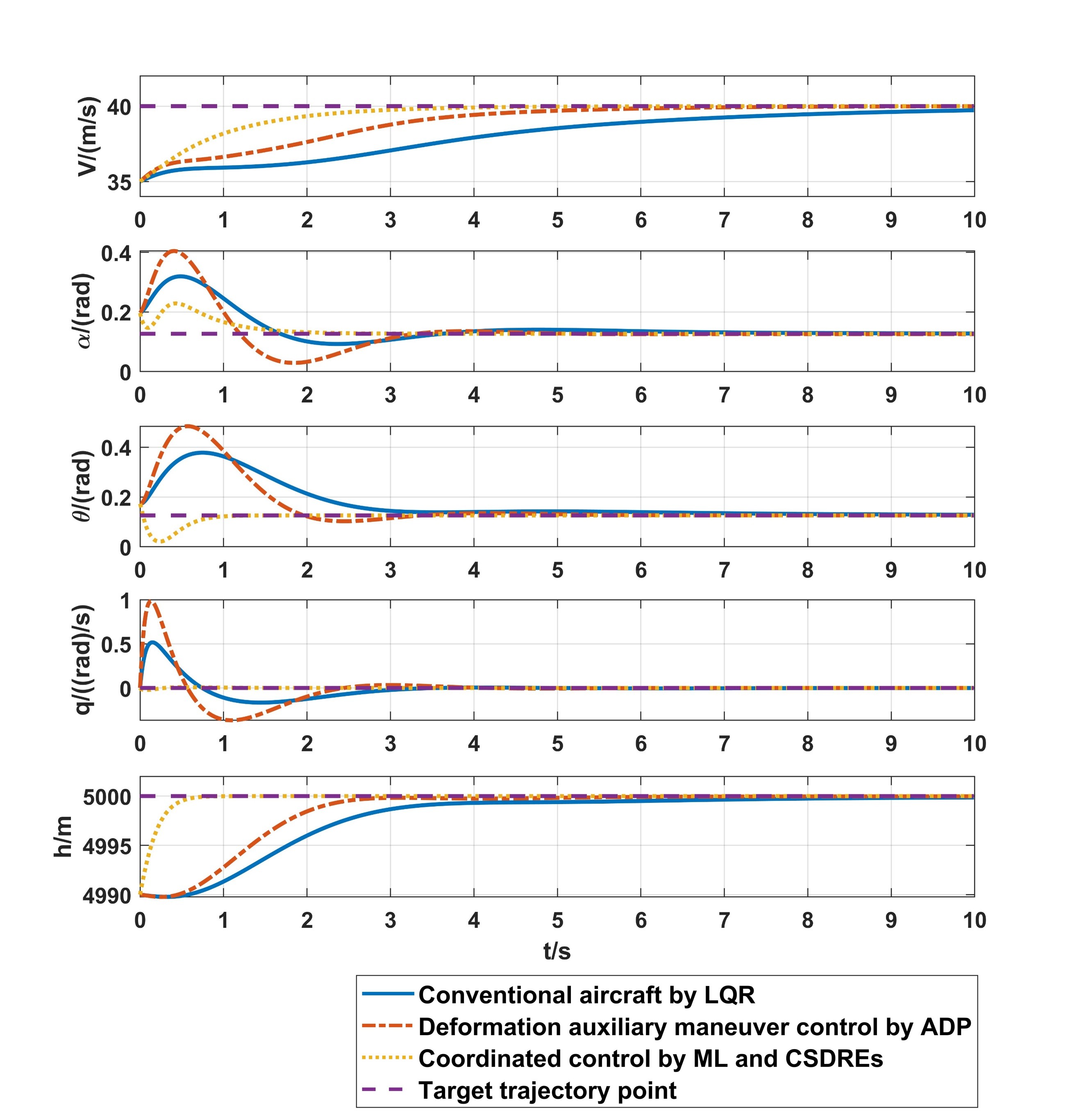}
			\caption{{\small$\mathbf{The~curves~of~states~\bm x_n}$. This figure illustrates the evolution trajectories of flight velocity,    attack angle, pitch angle, pitch angular velocity
  and flight altitude
 for the MA driven by the proposed coordinated control strategies, conventional aircraft by LQR and deformation auxiliary maneuver control by ADP in \cite{greene2023deep}.}}
			\label{fig1}
\vspace{-10pt}
\end{figure}
		\vspace{-5mm}
	{\subsection{Comparisons with  Other Methods}
 To demonstrate the advantages of the proposed coordinated  control strategies, we compare it with several existing control algorithms, such as conventional aircraft based on the linear quadratic regulator (LQR) method and coordinated strategies based on the model-based approximate dynamic programming (ADP) method \cite{greene2023deep}.

In the design of conventional aircraft control, we assume that the morphing ratio is fixed at {\small$\xi_n=0.2$}. We establish the linear model of the system \eqref{equ:C7} through Jacobian linearization, i.e., {\small$\dot{\bm x} =\bm A_{*} \bm x +\bm B_{*} \bm u$}, where
\begin{equation*}
			\small\begin{split}
				&\bm A_{*}=\\&\begin{bmatrix*}[c]
					-0.0353&3.91 &-9.8&0&7.54\times 10^{-5}\\
					-0.0127&-1.43& 2.14\times 10^{-4}&0.95&2.9\times 10^{-5}\\
					0&0& 0&1&0\\
					-0.0614&-8.42&0&-1.72&1.32\times 10^{-4}\\
					-8.73\times10^{-4}&-40&40&0&0\\
				\end{bmatrix*},
\end{split}
\end{equation*}
\begin{equation*}
\small\begin{split}
\bm B_{*}=\begin{bmatrix}
					0&0.0169\\
					-0.0822&-5.4\times 10^{-5}\\
					0 & 0\\
					-4.2074&0\\
					0&0\\
				\end{bmatrix}.
\end{split}
\vspace*{-8.2mm}
\end{equation*}
The optimal performance index function is chosen as {\small$J=\d\frac{1}{2}\int_{0}^{\infty}\{ \bm{x}^{T}(t)\bm Q\bm{x}(t)+\bm{\bm{u}}^{T}(t)\bm{R}\bm{\bm{u}}(t) \}dt$}, in which {\small$\bm Q$} {\small$=
\rm{ diag}\{200,8000,8000,30000,200\}$} and {\small$\bm R$} {\small$=\rm {diag}\{3000,0.5\}$} for the LQR method. 
 
 This paper also compares the proposed method with the coordinated strategies generated by other control algorithm.  We integrate structural deformation with conventional manipulation into a new input vector, establishing a longitudinal nonlinear dynamic model for  deformation-assisted maneuvering of MA, represented  as {\small$\dot{\bm x}=\bm f_m(\bm x)+\bm g_m(\bm x)\bm u_m$} (\cite{yin2015coordinated}, Ch. 5), where {\small$\bm u_m=[\delta_e~~ \delta_t~~ \xi]^{T}$}. A model-based RL ADP framework in \cite{greene2023deep} is extended 
 to solve the infinite horizon optimal  problem of \eqref{equ:C37}.

Fig. 5 shows the comparative simulation results mentioned above. The above simulation results indicate that the suggested coordination strategies provide a better flight altitude and
 velocity tracking performance in comparison to the other two flight control algorithms. 
 As shown in Fig. 5, compared to conventional aircraft, the MA based on proposed coordinated method of deformation and flight can converge to equilibrium
points more rapidly. Additionally, due to the interaction between deformation control and flight control, the proposed  method can save {\small$40.26\%$} of cost compared to traditional aircraft using LQR. In comparison to deformation auxiliary
maneuver control by ADP, the MA based on the proposed method can converge to equilibrium
points more accurately  and can quickly adjust to new trajectories with almost no overshoot.
 Based on the 
cumulative cost, the proposed  coordinated 
method of deformation and flight improves the performance  by {\small$25.35\%$} compared with the deformation
auxiliary maneuver control by ADP. }
 \vspace*{-5mm}
\section{CONCLUSION}
		In this work,   an integrated approach has been presented to achieve coordinated control of deformation and flight for MA.
		The ML formulation for the offline training  assists in design of { a} common representation function  that can represent the unknown dynamics effectively.  Furthermore,  on the basis of learned {representation} function, a non-cooperative differential game
		is constructed to describe the coordinated control of deformation and flight for the MA. The feedback Nash equilibrium can be obtained by solving a pair of
		CSDREs. 
		The comparative analyses against existing methods demonstrates
that the proposed coordinated 
method of deformation and flight improves performance by {\small$40.26\%$} compared with the conventional aircraft by LQR, and improves performance by {\small$25.35\%$} compared with the deformation
auxiliary maneuver control by ADP.

		Nevertheless,  it is worth pointing out that the current study only provides a fundamental framework. The effects of noise and representation errors  have been disregarded in this paper. 
{In addition, the system targeted by the method in this paper is a nonlinear affine system, where wingspan deformation has no impact on flight control inputs. In future work, we will further consider systems where deformation affects flight control inputs \cite{oktay2017simultaneous,eraslan2023multidisciplinary}.}

{\small\begin{IEEEbiography}[{\includegraphics[width=1in,height=1.3in,clip,keepaspectratio]{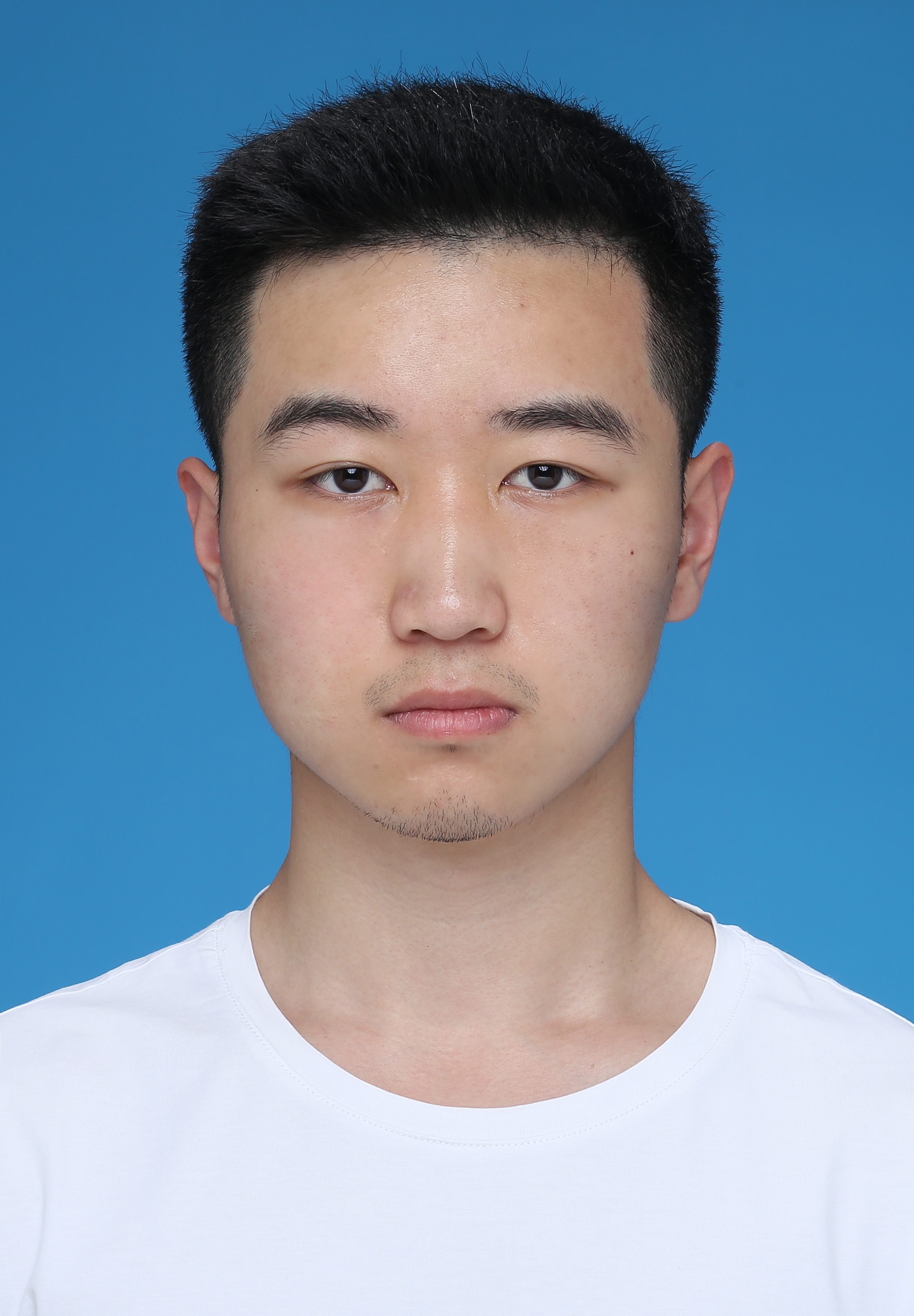}}]{Hao-Chi Che}received the B.E. degree in electrical engineering and automation from the Applied Technology College of Soochow University, Suzhou, China, in 2018, and the M.S. degree in control theory and control 
engineering from the School of Mechanical and Electrical Engineering, Soochow University, Suzhou, China, in 2021. He is currently pursuing the 
Ph.D. degree in control science and engineering with the School 
of Automation Science and Electrical Engineering, Beihang 
University, Beijing, China. 

His current research interests include game control, machine learning and neural network-based control. 
\end{IEEEbiography}}

\begin{IEEEbiography}[{\includegraphics[width=1in,height=1.2in,clip,keepaspectratio]{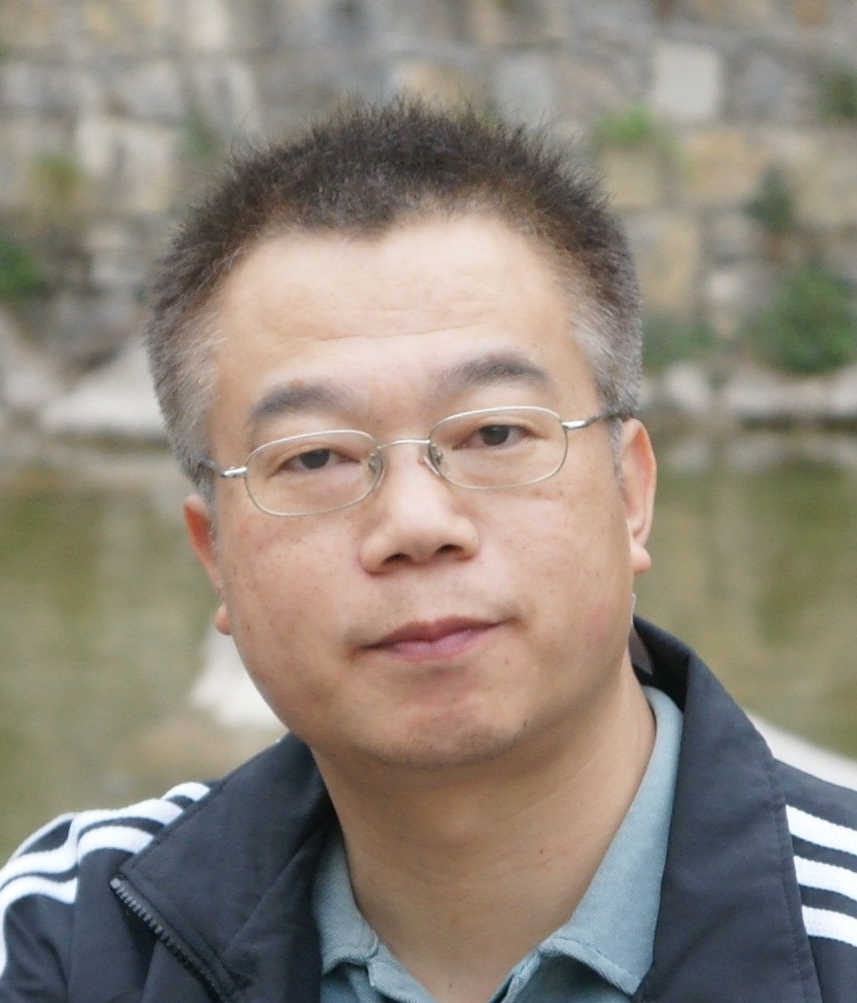}}]{Huai-Ning Wu} received the B.E. degree in automation from 
Shandong Institute of Building Materials Industry, Jinan, China, 
and the Ph.D. degree in control theory and control engineering 
from Xi’an Jiaotong University, Xi’an, China, in 1992 and 1997, 
respectively.

From August 1997 to July 1999, he was a Post-Doctoral Research Fellow with the Department of
 Electronic Engineering, Beijing Institute of Tech
nology, Beijing, China. Since August 1999, he has
 been with the School of Automation Science and
 Electrical Engineering, Beihang University (formerly Beijing University of
 Aeronautics and Astronautics), Beijing, China. From December 
2005 to May 2006, he was a Senior Research Associate with the City University of Hong Kong (CityU), Hong Kong. From October 2006 to December 2008 and 
from July 2010 to August 2013, he was a Research Fellow with CityU. He is 
currently a Professor with Beihang University. His current research interests 
include robust and fault-tolerant control, intelligent control, distributed parameter 
systems, and human-in-the-loop systems.
 
 Dr. Wu  serves as an Associate Editor for the IEEE
 Transactions on Fuzzy Systems.
\end{IEEEbiography}


\vspace*{-5mm}
		{\small\begin{thebibliography}{[20d]}
			\setcounter{enumiv}{0}
			\bibitem{Bonnema1988AFTIF111MA}
			K.~L. Bonnema and S.~W. Smith, ``{AFTI/F-111} mission adaptive wing flight
			research program,'' in \emph{4th Flight Test Conference}, 2006, pp. 155--161.
			
			\bibitem{Pendleton2000ActiveAW}
			E.~W. Pendleton, D.~Bessette, P.~B. Field \emph{et~al.}, ``Active aeroelastic
			wing flight research program: Technical program and model analytical
			development,'' \emph{J. Aircraft}, vol.~37, no.~4, pp. 554--561,
			2000.
			
			\bibitem{Kudva2004OverviewOT}
			Y. Jia, H. Chen, W. Q. Wang \emph{et~al.}, ``Robust adaptive beamforming based on manifold analysis for flexible conformal array of smart morphing wing aircraft," \emph{ IEEE T. Aero. Elec. Sys.}, vol. 60, no. 4, pp. 4753-4766,  2024.
			
			\bibitem{Seigler2007ModelingAF}
			T.~M. Seigler, D.~A. Neal, J.~Bae \emph{et~al.}, ``Modeling and flight control
			of large-scale morphing aircraft,'' \emph{J. Aircraft}, vol.~44,
			no.~4, pp. 1077--1087, 2007.
			
			\bibitem{boutayeb1999strong}
			M.~Boutayeb and D.~Aubry, ``A strong tracking extended {Kalman} observer for
			nonlinear discrete-time systems,'' \emph{IEEE T. Automat. Contr.}, vol.~44, no.~8, pp. 1550--1556, 1999.
			
%
			
			\bibitem{gong2019disturbance}
			L.~Gong, Q.~Wang, and C.~Dong, ``Disturbance rejection control of morphing
			aircraft based on switched nonlinear systems,'' \emph{Nonlinear Dynam.},
			vol.~96, pp. 975--995, 2019.
			
			\bibitem{wu2019new}
			K.~Wu, P.~Zhang, and H.~Wu, ``A new control design for a morphing {UAV} based
			on disturbance observer and command filtered backstepping techniques,''
			\emph{Sci. China Technol. Sc.}, vol.~62, pp. 1845--1853, 2019.
			
			\bibitem{JIANG20151640}
			Y.~Liang, L.~Zhang, and X.~Wang,  ``Anti-bump switched LPV control with delayed scheduling for morphing aircraft,'' \emph{IEEE T. Neur. Net. Lear.}, vol.~60, no.~4, pp. 5010-5023, 2024.
			
			\bibitem{lee2019linear}
			J.~Lee, S.~Kim, S.~Jung \emph{et~al.}, ``Linear parameter-varying control of
			variable span-sweep morphing aircraft,'' in \emph{AIAA Scitech 2019 Forum},
			2019, p. 0106.
			
			\bibitem{YUE2013909}
 G.~Gai, T.~Wu, M.~Hao, et al. ``Dynamic event-triggered gain-scheduled {$H_\infty$} control for a polytopic {LPV} model of morphing aircraft,'' \emph{IEEE T. Aero. Elec. Sys.}, vol.~61, no.~1, pp. 93-106.

			
			\bibitem{seigler2009analysis}
			T.~Seigler and D.~Neal, ``Analysis of transition stability for morphing
			aircraft,'' \emph{J. Guid. Control Dynam.}, vol.~32,
			no.~6, pp. 1947--1954, 2009.
			
			\bibitem{shi2015morphing}
			R.~Shi and J.~Peng, ``Morphing strategy design for variable-wing aircraft,'' in
			\emph{15th AIAA Aviation Technology, Integration, and Operations Conference},
			2015, p. 3002.
			
			\bibitem{gandhi2009hardware}
			N.~Gandhi, J.~Cooper, D.~Ward \emph{et~al.}, ``A hardware demonstration of an
			integrated adaptive wing shape and flight control law for morphing
			aircraft,'' in \emph{AIAA Guidance, Navigation, and Control Conference},
			2009, p. 5890.
			
			\bibitem{guo2018compound}
			J.~Guo, L.~Wu, and J.~Zhou, ``Compound control system design for asymmetric
			morphing wing aircraft,'' \emph{Journal of Astronautics}, vol.~39, no.~1,
			pp.~52-59, 2018.
			
			\bibitem{yin2015coordinated}
			M.~Yin, ``Coordinated control of deformation and flight for morphing
			aircraft,'' \emph{Nanjing University of Aeronautics and Astronautics}, 2015.
			
		
			


           \bibitem{liu2020optimal}
			J.~Liu, J.~Shan, Y.~Hu \emph{et~al.}, ``Optimal switching control for morphing
			aircraft with aerodynamic uncertainty,'' in \emph{IEEE 16th International
				Conference on Control \& Automation}.\hskip 1em plus 0.5em minus 0.4em\relax
			IEEE, 2020, pp. 1167--1172.
			
%
%
			
			\bibitem{lecun2015deep}
			Y.~LeCun, Y.~Bengio, and G.~Hinton, ``Deep learning,'' \emph{Nature}, vol. 521,
			no. 7553, pp. 436--444, 2015.
			
			\bibitem{littman2015reinforcement}
			M.~L. Littman, ``Reinforcement learning improves behaviour from evaluative
			feedback,'' \emph{Nature}, vol. 521, no. 7553, pp. 445--451, 2015.
			
			\bibitem{mnih2015human}
			V.~Mnih, K.~Kavukcuoglu, D.~Silver \emph{et~al.}, ``Human-level control through
			deep reinforcement learning,'' \emph{Nature}, vol. 518, no. 7540, pp.
			529--533, 2015.
			
			\bibitem{lillicrap2015continuous}
			T.~P. Lillicrap, J.~J. Hunt, A.~Pritzel \emph{et~al.}, ``Continuous control
			with deep reinforcement learning,'' \emph{arXiv preprint arXiv:1509.02971},
			2015.
			
			\bibitem{schulman2017proximal}
			J.~Schulman, F.~Wolski, P.~Dhariwal \emph{et~al.}, ``Proximal policy
			optimization algorithms,'' \emph{arXiv preprint arXiv:1707.06347}, 2017.
			
				\bibitem{valasek2005reinforcement}
			J.~Valasek, M.~D. Tandale, and J.~Rong, ``A reinforcement learning-adaptive
			control architecture for morphing,'' \emph{Journal of Aerospace Computing,
				Information, and Communication}, vol.~2, no.~4, pp. 174--195, 2005.

\bibitem{wen2017deep}
			 N.~Wen, Z.~Liu, and L.~Zhu, `Deep reinforcement learning and its application on autonomous shape optimization for morphing aircrafts,'' \emph{Journal of Astronautics}, vol.~38, no.~11, pp. 1153--1159, 2017.
			
			\bibitem{li2020morphing}
			R.~Li, Q.~Wang, Y.~Liu \emph{et~al.}, ``Morphing strategy design for {UAV}
			based on prioritized sweeping reinforcement learning,'' in \emph{IECON 2020
				The 46th Annual Conference of the IEEE Industrial Electronics Society}.\hskip
			1em plus 0.5em minus 0.4em\relax IEEE, 2020, pp. 2786--2791.
			
			\bibitem{snell2017prototypical}
			J.~Snell, K.~Swersky, and R.~Zemel, ``Prototypical networks for few-shot
			learning,'' \emph{Advances in Neural Information Processing Systems},
			vol.~30, 2017.
			
			\bibitem{kirkpatrick2017overcoming}
			J.~Kirkpatrick, R.~Pascanu, N.~Rabinowitz \emph{et~al.}, ``Overcoming
			catastrophic forgetting in neural networks,'' \emph{P.
				Natl. Acad. Sci. USA}, vol. 114, no.~13, pp. 3521--3526, 2017.
			
			\bibitem{zenke2017continual}
			F.~Zenke, B.~Poole, and S.~Ganguli, ``Continual learning through synaptic
			intelligence,'' in \emph{International Conference on Machine Learning}.\hskip
			1em plus 0.5em minus 0.4em\relax PMLR, 2017, pp. 3987--3995.
			
			\bibitem{santoro2016meta}
			A.~Santoro, S.~Bartunov, M.~Botvinick \emph{et~al.}, ``Meta-learning with
			memory-augmented neural networks,'' in \emph{International Conference on
				Machine Learning}.\hskip 1em plus 0.5em minus 0.4em\relax PMLR, 2016, pp.
			1842--1850.
			
			\bibitem{finn2017model}
			C.~Finn, P.~Abbeel, and S.~Levine, ``Model-agnostic meta-learning for fast
			adaptation of deep networks,'' in \emph{{International Conference on Machine
					Learning}}.\hskip 1em plus 0.5em minus 0.4em\relax PMLR, 2017, pp.
			1126--1135.
			
			\bibitem{hospedales2021meta}
			T.~Hospedales, A.~Antoniou, P.~Micaelli \emph{et~al.}, ``Meta-learning in
			neural networks: A survey,'' \emph{IEEE T. Pattern Anal.}, vol.~44, no.~9, pp. 5149--5169, 2021.
			
			\bibitem{cloutier1996nonlinear}
			J.~R. Cloutier, C.~N. D'Souza, and C.~P. Mracek, ``Nonlinear regulation and
			nonlinear {$H_\infty$} control via the state-dependent {Riccati} equation
			technique: Part 1, theory,'' in \emph{Proceedings of the International
				Conference on Nonlinear Problems in Aviation and Aerospace}, 1996, pp.
			117--131.
			
			\bibitem{cloutier1997state}
			J.~R. Cloutier, ``State-dependent {Riccati} equation techniques: an overview,''
			in \emph{Proceedings of the 1997 American Control Conference}, vol.~2.\hskip
			1em plus 0.5em minus 0.4em\relax IEEE, 1997, pp. 932--936.
			
			\bibitem{dutka2005optimized}
			A.~S. Dutka, A.~W. Ordys, and M.~J. Grimble, ``Optimized discrete-time state
			dependent {Riccati} equation regulator,'' in \emph{Proceedings of the 2005
				American Control Conference}.\hskip 1em plus 0.5em minus 0.4em\relax IEEE,
			2005, pp. 2293--2298.
			
			\bibitem{2012Survey}
			T.~{\c{C}}imen, ``Survey of state-dependent {Riccati} equation in nonlinear
			optimal feedback control synthesis,'' \emph{J. Guid. Control Dynam.}, vol.~35, no.~4, pp. 1025--1047, 2012.
			
			\bibitem{wang2018coupled}
			X.~Wang, E.~E. Yaz, and S.~C. Schneider, ``Coupled state-dependent {Riccati}
			equation control for continuous time nonlinear mechatronics systems,''
			 \emph{J. Dyn. Syst-T. ASME}, vol. 140, no. 11, p. 111013, 2018.
			
			\bibitem{Yang1983}
			B.~Yang, ``Formula expression of standard atmospheric parameters,''
			\emph{Journal of Astronautics}, vol.~1, pp. 86--89, 1986.
			
			\bibitem{o2022neural}
			M.~O’Connell, G.~Shi, X.~Shi \emph{et~al.}, ``Neural-fly enables rapid
			learning for agile flight in strong winds,'' \emph{Sci. Robot.}, vol.~7,
			no.~66, p. eabm6597, 2022.
			
			\bibitem{shi2019neural}
			G.~Shi, X.~Shi, M.~O’Connell \emph{et~al.}, ``Neural lander: Stable drone
			landing control using learned dynamics,'' in \emph{2019 International
				Conference on Robotics and Automation}.\hskip 1em plus 0.5em minus
			0.4em\relax IEEE, 2019, pp. 9784--9790.
			
			\bibitem{bartlett2017spectrally}
			P.~L. Bartlett, D.~J. Foster, and M.~J. Telgarsky, ``Spectrally-normalized
			margin bounds for neural networks,'' \emph{Advances in Neural Information
				Processing Systems}, vol.~30, 2017.
			
			\bibitem{Li1995LyapunovIF}
			T.~Li and Z.~Gajic, ``Lyapunov iterations for solving coupled algebraic
			{Riccati} equations of nash differential games and algebraic {Riccati}
			equations of zero-sum games,'' in \emph{New Trends in Dynamic Games and
				Applications}.\hskip 1em plus 0.5em minus 0.4em\relax Springer, 1995, pp.
			333--351.
			
			\bibitem{Berger1968PerspectivesIN}
			M.~S. Berger and G.~Berger, \emph{Perspectives in nonlinearity : an
				introduction to nonlinear analysis}.\hskip 1em plus 0.5em minus 0.4em\relax
			W.A.Benjamin, Inc. New York, 1968.
			


			\bibitem{Kantorovich1952FunctionalAI}
			L.~V. Kantorovich and G.~P. Akilov, \emph{Functional analysis in normed
				spaces}.\hskip 1em plus 0.5em minus 0.4em\relax New York: Mamillan, 1964.

{\bibitem{LinPayload}
		 D.~Lin, J. Han, K. Li, \emph{et~al.}, ``Payload transporting with two quadrotors by centralized reinforcement learning method,'' \emph{IEEE T. Aero. Elec. Sys.}, vol. 60, no.1, pp. 239-251, 2023.}

{\bibitem{good2014}
			I.~Goodfellow, J.~Pouget-Abadie,  M.~Mirza \emph{et~al.}, ``Generative adversarial nets,'' \emph{ Advances in Neural Information Processing Systems}, vol. 27, 2014.

\bibitem{greene2023deep}
			M.~L.~Greene, Z.~ I.~ Bell, S.~Nivison \emph{et~al.}, ``Deep neural network-based approximate optimal tracking for unknown nonlinear systems,'' \emph{IEEE T. Automat. Contr.}, vol.~68, no.~5, pp. 3171--3177, 2023.
          
           \bibitem{oktay2017simultaneous}
			T.~Oktay and S.~Coban, ``Simultaneous longitudinal and lateral flight control systems design for both passive and active morphing TUAVs,'' \emph{Elektron. Elektrotech.}, vol.~23, no.~5, pp. 15--20, 2017.

           \bibitem{eraslan2023multidisciplinary}
			Y.~Eraslan and T.~Otkay, ``Multidisciplinary performance enhancement on a fixed-wing unmanned aerial vehicle via simultaneous morphing wing and control system design,'' \emph{Inf. Technol. Control}, vol.~52, no.~4, pp. 833--848, 2023.}
\end{thebibliography}}
\end{document}